\documentclass[aps,prd,superscriptaddress,nofootinbib,amsmath,amsfonts,preprintnumbers,groupedaddress,10pt,english]{revtex4}
\usepackage{amsmath}
\usepackage{amssymb}
\usepackage{babel}
\usepackage{wrapfig}
\usepackage{cancel}

\usepackage{relsize,exscale}
\makeatletter

\usepackage{array,multirow,graphicx}
\usepackage{dcolumn}
\usepackage{newlfont}
\usepackage{bm}
\usepackage[colorlinks,citecolor=blue,urlcolor=blue,linkcolor=blue]{hyperref}
\usepackage[figtopcap]{subfigure}
\usepackage{color}
\usepackage{verbatim}

\usepackage{scalerel}
\usepackage{tikz}
\usetikzlibrary{svg.path}
\definecolor{orcidlogocol}{HTML}{A6CE39}
\tikzset{
  orcidlogo/.pic={
    \fill[orcidlogocol] svg{M256,128c0,70.7-57.3,128-128,128C57.3,256,0,198.7,0,128C0,57.3,57.3,0,128,0C198.7,0,256,57.3,256,128z};
    \fill[white] svg{M86.3,186.2H70.9V79.1h15.4v48.4V186.2z}
                 svg{M108.9,79.1h41.6c39.6,0,57,28.3,57,53.6c0,27.5-21.5,53.6-56.8,53.6h-41.8V79.1z M124.3,172.4h24.5c34.9,0,42.9-26.5,42.9-39.7c0-21.5-13.7-39.7-43.7-39.7h-23.7V172.4z}
                 svg{M88.7,56.8c0,5.5-4.5,10.1-10.1,10.1c-5.6,0-10.1-4.6-10.1-10.1c0-5.6,4.5-10.1,10.1-10.1C84.2,46.7,88.7,51.3,88.7,56.8z};}}
\newcommand\orcid[1]{\href{https://orcid.org/#1}{\mbox{\scalerel*{
\begin{tikzpicture}[yscale=-1,transform shape]
\pic{orcidlogo};
\end{tikzpicture}
}{|}}}}
\graphicspath{{./}{Figs/}}
\begin{document}
\date{\today}

\title{ Stability of a realistic astrophysical pulsar and its mass-radius relation in higher-order curvature gravity }

\author{G.~G.~L.~Nashed~\orcid{0000-0001-5544-1119}}
\email{nashed@bue.edu.eg}
\affiliation {Centre for Theoretical Physics, The British University, P.O. Box
43, El Sherouk City, Cairo 11837, Egypt}

\author{Kazuharu Bamba~\orcid{0000-0001-9720-8817}}
\email{bamba@sss.fukushima-u.ac.jp}
\affiliation {Faculty of Symbiotic Systems Science,
Fukushima University, Fukushima 960-1296, Japan}




\begin{abstract}
The objective of this research is to explore compact celestial objects while considering the framework of an extended gravitational theory known as $\mathcal{R}+f(\mathcal{G})$ gravity.  The notations $\mathcal{R}$ and $\mathcal{G}$ denote the Ricci scalar and the Gauss-Bonnet invariant, respectively. Radio pulsars, which are neutron stars with masses greater than 1.8 times that of the Sun ($M_\odot$), provide exceptional opportunities for delving into fundamental physics in extraordinary environments unparalleled in the observable universe and surpassing the capabilities of experiments conducted on Earth.   Through the utilization of both the linear and quadratic expressions of the function { $f(\mathcal{G}) = \alpha_1 \mathcal{G}^2$, where  $\alpha_1$ (with dimensional units of [${\textit length}^6$]) are incorporated}, we have achieved an accurate analytical solution for anisotropic perfect-fluid spheres in a state of hydrostatic equilibrium.   By integrating the dimensional parameters $\alpha_1$ and the compactness factor, defined as ${\mathcal C=\frac{2GM}{Rc^2}}$, we showcase our capacity to encompass and depict all physical characteristics within the stellar structure.  We illustrate that the model can produce a stable arrangement encompassing its physical and geometric properties.   We illustrate that by utilizing the quadratic form of $\mathcal{G}$ in the $\mathcal{R}+f(\mathcal{G})$ framework, the ansatz of Krori-Barua establishes connections between pressure in the radial direction ($p_r$) using semi-analytical methods, pressure in the tangential direction ($p_t$), and density ($\rho$).  { Remarkably, in the context of the positive/negative $\alpha_1$ quadratic $\mathcal{R}+f(\mathcal{G})$ gravity framework, the maximum compactness is inherently restricted to values that fail below the Buchdahl limit when the surface density exceeds $4\times 10^{14}$ g/cm$^3$. In contrast to General Relativity (GR), where compactness is unlikely to approach the black hole limit, our study reveals a different trend.
We draw the relation between mass and radius associated with the boundary density we have determined, and this diagram is consistent with other observational discoveries.}
\end{abstract}

\maketitle
\section{Introduction}\label{Sec:Introduction}

{ Anisotropic pressure has been imposed in many models developed to derive realistic stellar models within the Einstein general relativity context, and in modified gravity as well. These models are motivated by the claim that pressure at the core of the compact star model could have an anisotropic structure where the density exceeds the nuclear density $10^{15} g/cm^3$. Herrera and Santos \cite{Herrera:1997plx} have looked into how things can be different in different directions, called local anisotropy, and how it influences whether self-gravitating systems stay stable or collapse under gravity. Other researchers have also explored this topic in related studies; see for example, \cite{Herrera:2007kz}.  More and more people are getting interested in anisotropic fluids, which are fluids where the pressure pushing outwards isn't the same in all directions. Recent studies suggest that forcing these fluids to be the same in all directions, known as enforcing local isotropy, might not be the best approach. It could make it harder to accurately model objects influenced by their gravity. In compact objects like stars, many different things are happening that can cause pressure to be different in different directions. So, it's really important to figure out ways to accurately describe these situations where pressure isn't the same everywhere. Thus, investigating possible techniques for obtaining precise solutions that characterize anisotropic fluid distributions is crucial. Readers are referred to \cite{Herrera:2020gdg} for a thorough examination of the causes and effects of local anisotropy in astrophysical objects.}

The configuration and characteristics of compact celestial entities pose a fundamental challenge within the realm of general relativity.  This area of research was inaugurated as early as 1916 by Schwarzschild, who derived the inner solution for a spherical object with a uniform density, featuring zero pressure at its vacuum boundary \cite{schwarzschild1916gravitationsfeld}.  Although theoretically simple and of limited relevance to real stellar objects, the Schwarzschild interior solution garnered significant attention, leading to intensive examination of its properties and extensions \cite{florides1974new,gron1986charged,Kohri:2001jx,Gabbanelli:2019txr,Posada:2018agb,Calmet:2020tlj}. A pivotal milestone in the advancement of relativistic astrophysics can be attributed to the contributions of Tolman  \cite{Tolman:1934za,tolman1939static}   and Oppenheimer and Volkoff \cite{oppenheimer1939massive} (TOV), who derived the structural equations for compact, spherically symmetric general relativistic objects in a static configuration.  

Both theoretical and numerical explorations of the TOV equation culminated in the determination of the maximum mass limit for neutron stars, which was identified to be approximately $3.2M_{\odot}$ \cite{Rhoades:1974fn}.  This outcome was achieved through the application of causality principles, the highly rigid equation of state characterized by $p=\rho c^2$, and Le Chatelier's principle. Importantly, this result remains true when the equation of state (EoS)  is not fully understood within a certain range of densities. Conversely, Chandrasekhar \cite{Chandrasekhar:1931ftj}  derived a limiting mass of approximately $1.4M_{\odot}$ for white dwarfs. Both theoretical reasoning and observational findings subsequently gave rise to the widely accepted notion, upheld for a considerable duration, that the mass distribution of neutron stars is concentrated around a value roughly in the vicinity of $1.4M_{\odot}$ \cite{1983bhwd.book.....S}. This mass value is a consequence of the understanding that neutron stars rely on the dominant pressure exerted by neutron degeneracy following the collapse of a white dwarf. For a neutron star with a mass of approximately $1.4M_{\odot}$, the associated radius is expected to be around $10\approx 15$km, with average $\rho$ to be roughly $6\times 10^{14} {\mathrm g/cm^3}$.

Nevertheless, the conventional perspective on the masses of neutron stars has undergone a significant transformation in recent times, as more precise determinations of neutron star masses have become increasingly accessible \cite{Horvath:2021ang}. Extensive astronomical observations and the detection of gravitational waves have definitively shown that the masses of neutron stars span a significantly wider range than what one would expect solely based on the Chandrasekhar limit.  In their study, Margalit and Metzger \cite{Margalit:2017dij} combined electromagnetic (EM) and gravitational wave (GW) data from the binary neutron star (NS) merger GW170817 to constrain the radii and maximum mass $M_{max}$  of neutron stars. They established an upper limit of Mmax?=2.17M?? (90\% confidence), which is arguably less model-dependent than other current constraints.  This limitation is more stringent and less reliant on specific models compared to alternative constraints. The study conducted by Shibata et al.  \cite{Shibata:2017xdx} based on the same event indicated that the equation of state for neutron matter must exhibit significant rigidity, suggesting that $M_{max}$ of neutron stars should considerably exceed $2M_{\odot}$. This mass threshold is essential to allow for the formation of a stable, long-lasting massive neutron star as the outcome of the merger in binary systems like GW170817, where the initial combined mass overrides $2.73M_{\odot}$. Furthermore, considering the absence of a relativistic optical counterpart, we can deduce $M_{max}$ value for the neutron star in the range of approximately $2.15-2.25M_{\odot}$. Comparable estimates for the neutron star's maximum mass have been derived in the studies conducted by Ruiz and Rezzolla \cite{Ruiz:2017due,Rezzolla:2017aly}, respectively.  Using the Shapiro delay, alternative methods for estimating the masses of neutron stars produced results of approximately  $1.928\pm 0.017M_{\odot}$  for the pulsar PSR J1614-2230 1, and a mass of $2.14_{-0.09}^{+0.10}M_{\odot}$ for the Millisecond Pulsar MSPJ0740+6620 \cite{NANOGrav:2019jur}.

One significant pathway for potentially elucidating the elevated mass values of neutron stars involves the realm of modified gravity. Modified theories have primarily appeared to account for contemporary expansion of the cosmos. To delve into in-depth conversations concerning the accelerating Universe, the concept of dark energy, and matters pertaining to modified gravity, consider consulting the reviews listed in the provided references \cite{Nojiri:2010wj,Joyce:2014kja,Nojiri:2017ncd,Frusciante:2019xia}.

Modified theories of gravity can potentially provide fresh perspectives on comprehending the composition of compact celestial bodies. Several essential characteristics of star structures, including maximum masses, moments of inertia, and mass-radius relationships, differ from standard general relativity within modified gravitational theories \cite{Olmo:2019flu}.
 These alterations stem from the modification of the Poisson equation for gravitational potential, resulting in changes to the radius, central density, mass,  and luminosity of the stars \cite{Olmo:2019flu}.
  The measured masses of many neutron stars, which fall in the range of approximately  $2M_{\odot}$, have led to disagreements with predictions from realistic equations of state for dense matter. As a result, certain softer equations, especially those incorporating hyperons, have been ruled out \cite{Olmo:2019flu}. Advanced gravity theories that extend beyond general relativity have the potential to mitigate or resolve these discrepancies by introducing corrections into the generalized hydrostatic equilibrium equations of extended stellar models \cite{Olmo:2019flu}.

Different forms of general relativity modifications have been put forth to explain the observed properties of compact celestial bodies, such as neutron stars, and their mass distribution. These extensions encompass various gravity theories, including $f(\mathcal{G})$, $f(\mathcal{R},\mathcal{G})$, $f(\mathcal{R},T)$, and $f\left(\mathcal{R},L_m\right)$ types,
For a concise overview of the astrophysical consequences of the aforementioned theories, refer to (\cite{Olmo:2019flu} and  references therein). 

A captivating subset of modified gravity models, commonly denoted as $f(\mathcal{R},\mathcal{G})$ \cite{Lobo:2005vc,Rahaman:2012pg,Nojiri:2017ncd}, arises from the inspiration of string theory and is known as modified Gauss-Bonnet gravity. It provides an interesting way to efficiently generate the late-time acceleration phase.  Gravitation theories that combine the Gauss-Bonnet and Ricci scalars are found in the literature; studies of this type of theory can be found in \cite{Mehdizadeh:2015jra,Sharif:2014yea}. The modified Gauss-Bonnet theory holds a significant role in the field of astrophysics due to its potential to reshape our understanding of the universe and its fundamental structures. This theory, inspired by string theory and emerging from the broader landscape of modified gravity, offers a novel perspective on gravity's behavior in extreme astrophysical environments. It opens the door to exploring phenomena that cannot be explained by conventional general relativity such as the nature of compact objects like neutron stars and black holes, their mass distributions, and the equations of state for dense matter. Moreover, the modified Gauss-Bonnet theory provides a framework to investigate the late-time acceleration of the universe, a critical aspect of modern cosmology. Its flexibility in accommodating various forms of gravitational interactions and its ability to address long-standing astrophysical anomalies make it an essential tool for astrophysicists seeking to comprehend the cosmos in its entirety. This theory's potential to reconcile theory with observations and offer new avenues for astrophysical exploration makes it a prominent and vital component of contemporary astrophysical research. { The goal of this study is to leverage the higher-order aspects of Gauss-Bonnet gravity, drawn from its known properties and advantages. Gauss-Bonnet gravity finds frequent use in theoretical physics to extend general relativity by accommodating higher-order curvature corrections. This extension can provide insights into various phenomena such as black hole physics, cosmology, and gravitational wave physics, thereby offering a more comprehensive understanding of gravity beyond standard Einsteinian theory. Hence, the implicit motivation for employing Gauss-Bonnet gravity could be to investigate its implications within the astrophysical framework, determining its potential to yield a consistent stellar model. This aspect will be explored throughout the study.}

The study's structure includes a review in Section \ref{Sec:fG_gravity} for the framework $\mathcal{R}+f(\mathcal{G})$ modified theory. Section \ref{Sec:Model} applies quadratic $\mathcal{R}+f(\mathcal{G})=\mathcal{R}+\alpha_1\mathcal{G}^2$ gravity to the KB stellar model, ensuring compatibility with the Schwarzschild exterior vacuum solution through imposed matching conditions.  In Section \ref{Sec:Stability}, We use astrophysical data to study radius and mass of the pulsar ${\mathcal J0740+6620}$ to limit the dimensionless  $\alpha_1$.  Additionally, we assess the model's validity by applying diverse stability criteria. In Section \ref{Sec:EoS_MR}, we derive the Equations of State (EoS) governing the matter sector. We also analyze the associated mass-radius diagram and present the study's conclusion in Section \ref{Sec:Conclusion}.
\section{Distribution of anisotropic matter in $\mathcal{R}+f(\mathcal{G})$ gravity}\label{Sec:fG_gravity}

The overarching action for the modified $f(\mathcal{G})$ is as follows \cite{Shamir:2017rjz},
\begin{equation}\label{action}
\mathcal{L}= \frac{1}{2{\kappa}^{2}}\int d^{4}x
\sqrt{-g}[\mathcal{R}+f(\mathcal{G})]+\int
d^{4}x\sqrt{-g}\mathcal{L}_{M}\,.
\end{equation}
In this context,  $f(\mathcal{G})$ represents the arbitrary Gauss-Bonnet  (GB) function with   $\mathcal{G}$ being the GB, while $\kappa^2$ stands for the coupling constant.
The GB expression, $\mathcal{G}$, is defined as follows:
\begin{equation}
\mathcal{G}=\mathcal{R}^{2}-4\mathcal{R}_{\zeta\eta}\mathcal{R}^{\zeta\eta}+\mathcal{R}_{\zeta\eta\mu\nu}\mathcal{R}
^{\zeta\eta\mu\nu}\,.
\end{equation}
In this context,  $\mathcal{R}_{\zeta\eta}$ and $\mathcal{R}_{\zeta\eta\mu\nu}$ represent the  Ricci and Reimann tensors, respectively. When we vary Eq.~(\ref{action}) concerning $g_{\zeta\eta}$, it yields the following set of fourth-order non-linear field equations:
\begin{align}\label{fe}
&G_{\zeta\eta}-[2\mathcal{R}g_{\zeta\eta}\nabla^{2}+
2\mathcal{R}\nabla_{\zeta}\nabla_{\eta}+4g_{\zeta\eta}\mathcal{R}^{\mu\nu}\nabla_{\mu}\nabla_{\nu}+
4\mathcal{R}_{\zeta\eta}\nabla^{2}-4\mathcal{R}^{\mu}_{\zeta}\nabla_{\eta}\nabla_{\mu}-4\mathcal{R}^{\mu}_{\eta}
\nabla_{\zeta}\nabla_{\mu}-4\mathcal{R}_{\zeta\mu\eta\nu}\nabla^{\mu}\nabla^{\nu}]
f_{\mathcal{G}}\nonumber\\
&-
\frac{1}{2}g_{\zeta\eta}f+[2\mathcal{R}\mathcal{R}_{\zeta\eta}-4\mathcal{R}^{\mu}_{\zeta}\mathcal{R}_{\mu\eta}
-4\mathcal{R}_{\zeta\mu\eta\nu}R^{\mu\nu}+2\mathcal{R}^{\mu\nu\delta}_{\zeta}\mathcal{R}_{\eta\mu\nu\delta}]
f_{\mathcal{G}}=\kappa^{2}\mathrm{\textit{T}}_{\zeta\eta},
\end{align}
In this context, we have the d'Alembertian operator denoted as $\Box = \nabla^2 = \nabla_{\zeta}\nabla^{\zeta}$, the Einstein tensor ${G}_{\zeta\eta} = \mathcal{R}_{\zeta\eta} - \frac{1}{2}g_{\zeta\eta}\mathcal{R}$, the function $f$ represented as $f \equiv f(\mathcal{G})$, and $f_{\mathcal{G}}$ denoting the partial derivative of $f(\mathcal{G})$ concerning $\mathcal{G}$. The Einstein equations can be restored simply by setting $f(\mathcal{G}) = 0$.
Here  $\mathrm{\textit{T}}_{\zeta\eta}$ is the energy-momentum tensor   defined as:
\begin{equation}\label{emt}
\mathrm{\textit{T}}_{\zeta\eta}=-\frac{2}{\sqrt{-g}}\frac{\delta(\sqrt{-g}\mathcal{L}_{M})}{\delta g^{\zeta\eta}}.
\end{equation}
Furthermore, the energy-momentum tensor that depends on the metric can take the following shape:
\begin{equation}\label{emt1}
\mathrm{\textit{T}}_{\zeta\eta}=g_{\zeta\eta}\mathcal{L}_{M}-2\frac{\partial\mathcal{L}_{M}}{\partial
g^{\zeta\eta}}.
\end{equation}
The anisotropic for of    $T_{\zeta\eta}$ is defined as:
\begin{equation}\label{12}
{T^\zeta}_{\eta}=(\rho+p_{t})V^{\zeta}V_{\eta}-p_{t}{\delta^\zeta}_{\eta}+(p_{r}-p_{t})\xi^{\zeta}\xi_{\eta}.
\end{equation}
 The energy-momentum tensor, ${T}{^\mu}{_\nu}$,  is defined as: \[{ {T}{^\mu}{_\nu}}={diag(-\rho c^2\, ,p_{r}\, ,p_t\, ,p_t)}.\]

The trace part of Eq.\eqref{fe} is given by
\begin{align}\nonumber
-\mathcal{R}+2\mathcal{G}f_{\mathcal{G}}(\mathcal{G})+2\mathcal{R}\nabla^{2}f_{\mathcal{G}}(\mathcal{G})-4R_{\alpha\beta}\nabla^{\alpha}\nabla^{\beta}f_{\mathcal{G}}(\mathcal{G})=\kappa^{2} \mathrm{\textit{T}}.
\end{align}
It is significant to mention that in $f(\mathcal{G})$, the standard conservation equation of
the stress-energy tensor is not satisfied, i.e., covariant derivative of $T_{\alpha\beta}$ is non-zero ($\nabla^{\alpha}T_{\alpha\beta}\neq0$). This fact can be justified by taking the covariant derivative of Eq.\eqref{fe}, which on combing with the Bianchi identity $\nabla^{\alpha} G_{\alpha\beta}$, gives
\begin{align}\label{a1m1}
\kappa^{2}\nabla^{\alpha}T_{\alpha\beta}=-\frac{1}{2}g_{\alpha\beta}\nabla^{\alpha}f(\mathcal{G}).
\end{align}
{
It can be easily seen from the above expression that in general, the conservation law does not hold for this gravitational theory. For the linearized $\mathcal{R}+f(\mathcal{G})$ gravity, the
corresponding field equations, as the linear results, are the same as those of the linearized $R$ gravity. In this limit, the gravitational coupling constant $\kappa^2$ does not differ from Einstein's $8\pi G/c^4$ \cite{Wu:2018jve}. }

\section{The anisotropic Gauss-Bonnet coupled with matter model}\label{Sec:Model}
{ In this section, we will seek an anisotropic solution within the framework of $R+ f\left( \mathcal{G} \right)$. To achieve this, we will employ the following  particular model, \begin{equation}\label{VM}
 f(\mathcal{G})=\alpha_1\mathcal{G}^2\, ,
\end{equation} where  $\alpha_1$ is a dimensional quantity that has the unit  ${\textit lengh}^6$.}
   In this investigation, we will introduce the following line element:
   \begin{align}\label{RG}
    {\mathrm   ds^2 = g_{\mu \nu} dx^\mu dx^\nu=-e^{\psi}c^2dt^2 + e^{\lambda}dr^2 + r^2(d\theta^2 + \sin^2\theta d\phi^2)}\,.
   \end{align}
  The metric functions, represented as $\psi$ and $\lambda$, are two functions that depend solely on the radial coordinate $r$. Consequently, we can calculate { $\sqrt{-g}=r^2 \sin \theta\,e^{(\psi + \lambda)/2}$}, as well as the four-velocity vector. $v^\mu=(ce^{-\psi/2}, 0, 0, 0)$.

   By using Eq. (\ref{RG}) for computation, we obtain the Ricci and Gauss-Bonnet scalars as follows:
\begin{align}\label{RaG}
&\mathcal{R}=\frac {\left(\lambda'{r}^{2} -4\,r \right) \psi'-2\,\psi''{r}^{2}- \psi'^{2}{r}^{2}  +4\, \lambda' r+4\,{e^{\lambda  }}-4}{2{r}^{2}{e^{\lambda  }}} \,,\nonumber\\
&\mathcal{G}=\frac {2{e^{-2\,\lambda  }} \left\{  \left[  \lambda'\left(e^{\lambda}-3 \right) - \left( {e^{\lambda  } }-1 \right) \psi' \right] \psi'  -\left( 2\,{ e^{\lambda }}-2 \right)  \psi'' \right\} }{{r}^{2}}\,,
\end{align}
where $'\equiv d/dr$ and $''\equiv d^2/dr^2$, and so on. Utilizing the Ricci and Gauss-Bonnet values provided by Eq. (\ref{RaG}) in the field equations (\ref{fe}) after applying the constraints (\ref{VM}), we obtain:

\begin{align}\label{Rho1}
&\rho=\frac{1}{c^2 { \kappa^2}{r}^{6}  {e^{4\lambda}}}\left( 32\,\alpha_1{r}^{2} \left( {e^{\lambda}}-1 \right) ^{2}\psi''''+32\,r\alpha_1 \, \left( r \left( {e^{\lambda }}-1 \right) \psi'+ \left( 7r-3r{e^{\lambda}}  \right) \psi'_1 -4{e^{\lambda}}+4 \right)  \left( {e^{\lambda}} -1 \right) \psi''' -16\alpha_1 \psi' {r}^{2} \left( { e^{\lambda}}-1 \right)\right.\nonumber\\
  & \left. \times \left( {e^{\lambda}}-3 \right) \lambda'''+24 \alpha_1\,{r}^{2} \left( {e^{\lambda}}-1 \right) ^{2 } a''^{2}-8\left( 8{r}^{2} \left( {e^{\lambda}}-\frac{5}2 \right) \left( {e^{\lambda}}-1 \right) \lambda''+{r}^{2} \left( {e^{\lambda} }-1 \right) ^{2} a'^{2 }+9 \left(  \left( {e^{\lambda}}-{\frac {19}{9}} \right) ra'_1\right.\right.\right.\nonumber\\
  & \left.\left.\left.+{\frac {16}{9}}{e ^{\lambda}}-{\frac {16}{9}} \right)  r \left( {e^{\lambda}}-1 \right) \psi' -11\, \left(  \left( { e^{\lambda}} \right) ^{2}-{\frac {64}{11}}\,{e^{\lambda}}+{\frac {61}{11}} \right) { r}^{2} \lambda'{}^{2}-28\, \left( {e^{\lambda }}-{\frac {17}{7}} \right) r \left( {e^{\lambda}}-1 \right) a'_1-24\, \left( {e^{\lambda}}-1 \right) ^{2} \right) \alpha_1\,\psi''\right.\nonumber\\
  & \left.-16\, \left( r \left( {e^{\lambda}}-1 \right)  \left( {e^{\lambda}}-2 \right) a' -7/2\,r \left( {\frac {45}{7}}+ \left( {e ^{\lambda}} \right) ^{2}-{\frac {48}{7}}\,{e^{\lambda}} \right)\lambda'-12+16 \,{e^{\lambda}}-4\, \left( {e^{\lambda}} \right) ^{2} \right)  \psi' r\alpha_1\,\lambda''-2\,\alpha_1\,{r}^{2} \left( {e^{\lambda}}-1 \right) ^{2} a'^{4}\right.\nonumber\\
  & \left.+4 \alpha_1 \lambda' {r}^{2} \left( {e^{\lambda}}-1 \right)  \left( {e^{\lambda}}-3 \right)  \psi'^{3}+22\,\alpha_1\, \left( {r}^{2} \left(  \left( { e^{\lambda}} \right) ^{2}-{\frac {54}{11}}{e ^{\lambda}}+{\frac {47}{11}} \right)  \lambda'^{2}+{\frac {40}{11}}\,r \left( {e ^{\lambda}}-1 \right)  \left( {e^{\lambda}}-{\frac {11}{5}} \right) a'_1+{\frac {48}{11}} \left( {e^{\lambda}}-1 \right) ^ {2} \right) \psi'^{2}\right.\nonumber\\
  & \left.+24 \,\psi'\alpha_1\, \left( \left( { e^{2\lambda}} -{\frac {35}{3}}\,{e^{\lambda}}+14 \right) {r}^{2} \lambda'^{2}+\frac{10}3 \left( {\frac {33}{5}}-{\frac {34}{5}}\,{e^{\lambda}}+ { e^{2\lambda}}  \right) ra'_1-16\,{e^{\lambda}}+4 { e^{2\lambda}}+12 \right)\psi' +{r}^{4} {e^{3\lambda}} \left( {e^{\lambda}}-1+ a'_1 r \right)  \right) ,
\end{align}
\begin{align}\label{pr1}
&p_r=\frac {1 }{{ \kappa^2} \left( {e^{\lambda}} \right) ^{4}{r}^{5}}\left\{8\,\alpha_1\,r \left( {e^{\lambda}} -1 \right) ^{2} a''^{2}-16\,r\alpha_1\,\psi' \left( {e^{\lambda}}-1 \right)  \left( {e^{\lambda}}-3 \right) \psi'''-8\,\psi'\left( r \left( {e^{\lambda}}-1 \right)  \left( {e^{\lambda}}-5 \right) a' -2\, \left( {e^{\lambda}}-3 \right)\right.\right. \nonumber\\
&\left.\left.\left( r \left( {e^{\lambda}}-3 \right)\lambda' +2\,{e^{\lambda}}-2 \right) \right) \alpha_1\,\psi'' +8\,r \alpha_1\,a'^{2} \left( {e^{\lambda}}-3 \right) ^{2}a''_1+2\,\alpha_1\,r \left( {e^{\lambda}}-1 \right) ^{2}a'^{4}+4\, \left( {e^{\lambda}}-3 \right) \left( r \left( {e^{\lambda}}-3 \right)\lambda' +4\,{e^{\lambda}}-4 \right) \alpha_1\,\psi^{3}\right. \nonumber\\
&\left.-6\, \left( {e^{\lambda}}-3 \right)  \left( r \left( {e^{\lambda}}-7 \right) \lambda'+\frac{8}3\,{e^{\lambda}}-8 \right)  \lambda' \alpha_1\,\psi'^{2}+{r}^{4}\psi' \left( {e^{\lambda}} \right) ^{3}-{r}^{3} \left( {e^{\lambda}} \right) ^ {3} \left( {e^{\lambda}}-1 \right)\right\},
\end{align}
\begin{align}\label{pt1}
&p_t=\frac{1}{4{ \kappa^2}\left( {e^{\lambda}} \right) ^{4}{r}^{5}} \left( 64\alpha_1{r}^{2} \psi'  \left( {e^{\lambda}}-1 \right) \psi'''' +96 r\alpha_1 \left( \frac{2}3r \left( {e^{\lambda}}-1 \right) \psi''+ \psi' \left( r \left( {e^{\lambda}}-1 \right) \psi'-\frac{8}3\left( {e^{\lambda}}-\frac{7}4 \right) ra'_1+\frac{8}3-\frac{8}3{ e^{\lambda}} \right)  \right) \psi''' \right. \nonumber\\
&\left.-32\,\alpha_1{r}^{2} \psi'^{2} \left( {e^{\lambda}} -3 \right) \lambda'''+32r\alpha_1\, \left( 4r \left( {e^{\lambda}}-1 \right) a' -\left(3 {e^{\lambda}}-7 \right) r\lambda' +5-6{e^{\lambda}}+\left( {e^{\lambda}} \right) ^{2} \right)  \psi''^{2}+ \left( 160\left( {\frac {13}{5}}-{e^{\lambda}} \right)  \psi' {r}^{2}\alpha_1 \lambda''\right.\right. \nonumber\\
&\left.\left.+32\,\alpha_1\,{r}^{2} \left( {e^{\lambda}}-1 \right)  \psi'^{3}-304\, \left(  \left( -{\frac {33}{19}}+{e^{\lambda}} \right) r\lambda'-\frac{2}{19}\, \left( {e^{\lambda}}-1 \right)  \left( {e^{\lambda}} -13 \right)  \right) r\alpha_1\, \psi'^{2}+304\,\alpha_1\, \left(  \left( {e^{\lambda}}-{\frac {73}{19}} \right) {r}^{2} \lambda'{}^{2}\right.\right. \right.\nonumber\\
&\left.\left.\left.-2/19\,r \left(  \left( {e^{\lambda}} \right) ^{2}-24\,{e^{\lambda}}+43 \right) \lambda' -{\frac {24}{19}}+{ \frac {24}{19}}\,{e^{\lambda}} \right) \psi'+2\,{r}^{5} \left( {e^{\lambda} } \right) ^{3} \right) \psi'' -48 \, \psi'^{2}r \left( \left( {e^{\lambda}}-7/3 \right) r\psi'-3\,r \left( {e^{\lambda}}-5 \right)\lambda' -\frac{8}3\,{e^{\lambda}}\right.\right. \nonumber\\
&\left.\left.+8 \right) \alpha_1\,\psi''_1-16\, \left( r \left( {e^{\lambda}}-2 \right) \lambda' -\frac{5}2+3\,{e^{\lambda}}-1/2\, \left( {e^{\lambda}} \right) ^{2} \right) r\alpha_1\, \psi'^ {4}+96\, \left( {r}^{2} \left( {e^{\lambda}}-\frac{10}3 \right)  \lambda'^{2}-\frac{1}6\, r \left( -20\,{e^{\lambda}}+31+ \left( {e^{\lambda}} \right)^{2} \right) \lambda'\right.\right. \nonumber\\
&\left.\left.\left.+2\,{e^{\lambda}}-2 \right) \alpha_1\, \psi'^{3}+ \left( -80\, \left( {e^{\lambda}}-{\frac {42}{5}} \right) {r}^{2 }\alpha_1\, \left( \lambda'\right) ^{3}+8\, \alpha_1\,r \left( 141+ \left( {e^{\lambda}} \right) ^ {2}-34\,{e^{\lambda}} \right)  \lambda'{}^{2}-192\, \left( {e^{\lambda}}-3 \right) \alpha_1\,\lambda'\right.\right.\right. \nonumber\\
&\left.\left.+{r}^{ 5} \left( {e^{\lambda}} \right) ^{3} \right) \psi'^{2}-{r}^{4} \left( {e^{\lambda}} \right) ^{3} \left( -2+ \lambda'r \right) a'-2\,{r}^{4} \lambda'  \left( {e^{\lambda}} \right) ^{3} \right)\,.
\end{align}
It's evident that t density and the components of the pressures are altered and become identical with the GR  solution when considering $\alpha_1=0$ (as discussed in \cite{Nashed:2020kjh} and \cite{Roupas:2020mvs}).
\subsection{Krori-Barua model}\label{Sec:KB}
Because Eqs.~(\ref{Rho1}), (\ref{pr1}), and (\ref{pt1}) form complex systems comprising three non-linear differential equations with five unknowns, namely, $\psi$, $\lambda$, $\rho$, $p_r$, and $p_t$ additional conditions are required to render these systems solvable.   There are various approaches to impose these two constraints, with one option being to adopt specific equations of state (EoS). However, this method may not be particularly advantageous since the aforementioned systems, namely, Eqs.~(\ref{Rho1}), (\ref{pr1}), and (\ref{pt1}), involve fourth-order differential equations.  { The most effective approach we can employ is to prescribe specific forms for the ansatz of $\psi$ and $\lambda$. In this study, we will utilize the KB metric potentials, as presented in \cite{Krori1975ASS}. Anisotropic
compact star models in the Krori-Barua spacetime in { GR} have been studied in
Refs. \cite{Varela:2010mf,Rahaman:2011cw,2013EPJC...73.2409K,Bhar:2014mta,Bhar:2014mba,2015Ap&SS.360...32B} and in modified theories of gravity in Refs. \cite{2015Ap&SS.359...57A, 2016Ap&SS.361....8Z,2016Ap&SS.361..342Z,Ilyas:2017bmb,Yousaf:2018jkb,Saha:2019msh,Nashed:2023uvk,Nashed:2023lqj,Nashed:2023pxd,Nashed:2022zyi,Nashed:2021pkc}.}
\begin{equation}\label{eq:KB}
   {  \psi(r)=n_0 (r/R_s)^2+n_1,\,  \qquad \lambda(r)=n_2 (r/R_s)^2}\,,
\end{equation}
with $R_s$ representing the radius surface of the pulsar, and  ${n_0, n_1, n_2}$ are parameters that will be fixed from through the junction conditions.
For the sake of unit adjustment, we assume:
 \begin{equation}\label{const}
\alpha_1=\alpha_2R_s{}^6.\end{equation}
Using Eqs.~\eqref{eq:KB}, \eqref{const}, in  Eqs.~(\ref{Rho1}), (\ref{pr1}) and (\ref{pt1}), we obtain the explicate forms of energy-density, and the components of pressure  as:
\begin{align}\label{sol}
&  \rho=\frac{1}{{ \kappa^2} {e^{{\frac {4n_2\,{r}^{2}}{{R_s}^{2}}}}} {R_s}^{2}{r}^{6}{c}^{2}}\left(4480\,\alpha_2{n_2}^{3}n_0\,{e^{{\frac {n_2{r}^{2}}{ {R_s}^{2}}}}}{r}^{6} -384\,\alpha_2\,{n_2}^{3}n_0\, {e^{ {\frac {2n_2\,{r}^{2}}{{R_s}^{2}}}}}{r}^{6}-1728\,\alpha_2\,{n_0}^{2}{n_2}^{2}{ e^{{\frac {n_2\,{r}^{2}}{{R_s}^{2}}}}}{r}^{6}+352\,\alpha_2{n_0}^{2}{n_2}^{2} {e^{{\frac {2n_2\,{r}^{2}}{{R_s}^{2}}}}}{r}^{6}\right. \nonumber\\
&\left.+64\alpha_2{n_0}^{3}n_2 {e^{{\frac {2n_2\,{r}^{2}}{{R_s}^{2}}} }} {r}^{6}-256\,\alpha_2\,{n_0}^{3}n_2\,{ e^{{\frac {n_2\,{r}^{2}}{{R_s}^{2}}}}}{r}^{6}+64\,\alpha_2\,{R_s}^{4}{n_0}^{2}{e^{{\frac {n_2\,{r}^{2}}{{R_s}^{2}}} }}{r}^{2}-32\,\alpha_2\,{R_s}^{4}{n_0}^{2} {e^{{ \frac {2n_2\,{r}^{2}}{{R_s}^{2}}}}}{r}^{2}-64\,\alpha_2\,{n_0}^{3} {e^{{\frac {2n_2\,{r}^{2}}{{R_s }^{2}}}}} {R_s}^{2}{r}^{4}\right. \nonumber\\
&\left.+128\,\alpha_2\,{n_0}^{3} {e^{{\frac {n_2\,{r}^{2}}{{R_s}^{2}}}}}{R_s}^{2}{r}^{4}+2560\,\alpha_2\,{n_2}^{2}n_0\,{R_s}^{2}{r}^{4}-64\,\alpha_2\, {n_0}^{2}n_2\,{R_s}^{2}{r}^{4}+1152\,\alpha_2\,{R_s}^{4}n_0\,n_2\,{r}^{2}-  {e^{{\frac {3n_2\,{r}^{2} }{{R_s}^{2}}}}} {R_s}^{2}{r}^{4}+  {e^{{\frac {4n_2\,{r}^{2}}{{R_s}^{2}}}}} {R_s}^{2}{r}^{4}\right. \nonumber\\
&\left.-5376\,\alpha_2\,{n_2}^{3}n_0\,{r}^{6}+1504\,\alpha_2\,{n_0 }^{2}{n_2}^{2}{r}^{6}+192\,\alpha_2\,{n_0}^{3}n_2\, {r}^{6}-32\,\alpha_2\,{R_s}^{4}{n_0}^{2}{r}^{2}-64\,\alpha_2\,{n_0}^{3}{R_s}^{2}{r}^{4}-768\,\alpha_2\,{R_s}^{6}n_0\,{ e^{{\frac {n_2\,{r}^{2}}{{R_s}^{2}}}}}\right. \nonumber\\
&\left.+384\,\alpha_2\,{R_s} ^{6}n_0\,  {e^{{\frac {2n_2\,{r}^{2}}{{R_s}^{2}}}}}+64\,\alpha_2\,{n_0}^{4}{e^{{\frac {n_2\,{r}^{2}}{{R_s}^{2}}}}}{r}^{6}-32\,\alpha_2\,{n_0}^{4}{e^{{\frac {2n_2\,{r}^{2}}{{R_s}^{2}}}}} { r}^{6}+384\,\alpha_2\,{R_s}^{6}n_0+2\,n_2\, { e^{{\frac {3n_2\,{r}^{2}}{{R_s}^{2}}}}} {r}^{6}-32 \,\alpha_2\,{n_0}^{4}{r}^{6}\right. \nonumber\\
&\left.+128\,\alpha_2\,{n_0}^{2 }n_2\,{e^{{\frac {n_2\,{r}^{2}}{{R_s}^{2}}}}}{R_s}^{2}{r}^ {4}-64\,\alpha_2\,{n_0}^{2}n_2\, {e^{{ \frac {2n_2\,{r}^{2}}{{R_s}^{2}}}}}{R_s}^{2}{r}^{4}-1664 \,\alpha_2\,{R_s}^{4}n_0\,n_2\,{e^{{\frac {n_2 \,{r}^{2}}{{R_s}^{2}}}}}{r}^{2}+512\,\alpha_2\,{R_s}^{4}n_0\,n_2\, {e^{{\frac {2n_2\,{r}^{2}}{{R_s}^{2}}}}}{r}^{2}\right. \nonumber\\
&\left.+512\,\alpha_2\,{n_2}^{2}n_0\,  {e^{{\frac {2n_2\,{r}^{2}}{{R_s}^{2}}}}} { R_s}^{2}{r}^{4}-2816\,\alpha_2\,{n_2}^{2}n_0\,{e^{{ \frac {n_2\,{r}^{2}}{{R_s}^{2}}}}}{R_s}^{2}{r}^{4} \right) \,,
\nonumber\\
&p_r=\frac{1}{{ \kappa^2}  {e^{{\frac {4n_2\,{r}^{2}}{{R}^{2}}}}} {R}^{2}{r}^{4}}\left( 2\,n_0\, {e^{{\frac {3n_2\,{r}^{2}}{{R }^{2}}}}} {r}^{4}-384\,\alpha_2\,{n_0}^{2}n_2\,{e^{{\frac {n_2\,{r}^{2}}{{R}^{2}}}}}{R}^{2}{r}^{2}+576 \,\alpha_2\,{n_0}^{2}n_2\,{R}^{2}{r}^{2}+64\,\alpha_2\,{n_0}^{2}n_2\,  {e^{{\frac {2n_2\,{r}^{2} }{{R}^{2}}}}}{R}^{2}{r}^{2}\right. \nonumber\\
&\left.+960\,\alpha_2\,{n_0} ^{2}{n_2}^{2}{e^{{\frac {n_2\,{r}^{2}}{{R}^{2}}}}}{r}^ {4}-2016\,\alpha_2\,{n_0}^{2}{n_2}^{2}{r}^{4}+576\,\alpha_2\,{n_0}^{3}n_2\,{r}^{4}-96\,\alpha_2\,{n_0}^ {2}{n_2}^{2} {e^{{\frac {2n_2\,{r}^{2}}{{R}^{2}} }}} {r}^{4}+64\,\alpha_2\,{n_0}^{3}n_2\, {e^{{\frac {2n_2\,{r}^{2}}{{R}^{2}}}}}{ r}^{4}\right. \nonumber\\
&\left.-384\,\alpha_2\,{n_0}^{3}n_2\,{e^{{\frac {n_2\,{r}^{2}}{{R}^{2}}}}}{r}^{4}-576\,\alpha_2\,{R}^{4}{n_0}^{2}{e^{{\frac {n_2\,{r}^{2}}{{R}^{2}}}}}+416\,\alpha_2\,{R}^{4}{n_0}^{2}-64\,\alpha_2\,{n_0}^{4}{ e^{{\frac {n_2\,{r}^{2}}{{R}^{2}}}}}{r}^{4}+32\,\alpha_2\,{n_0}^{4}{r}^{4}+160\,\alpha_2\,{R}^{4}{n_0}^{2}  {e^{{\frac {2n_2\,{r}^{2}}{{R}^{2}}}}}\right. \nonumber\\
&\left.+32\alpha_2\,{n_0}^{4} {e^{{\frac {2n_2{r }^{2}}{{R}^{2}}}}} {r}^{4}+64\alpha_2\,{n_0}^{3 } {e^{{\frac {2n_2{r}^{2}}{{R}^{2}}}}} {R}^{2}{r}^{2}+64\alpha_2\,{n_0}^{3}{R}^{2}{r}^{2}-128\alpha_2\,{n_0}^{3}{e^{{\frac {n_2{r}^{2}}{{R}^{2 }}}}}{R}^{2}{r}^{2}+{e^{{\frac {3n_2\,{r}^{2}}{{R}^{ 2}}}}} {R}^{2}{r}^{2}-{e^{{\frac {4n_2 {r}^{2}}{{R}^{2}}}}} {R}^{2}{r}^{2} \right)\,,\nonumber\\
& {  p_t}=\frac{1}{{ \kappa^2} {e^{{\frac {4n_2\,{r}^{2}}{{R} ^{2}}}}} {R}^{4}{r}^{4}}\left( -n_2\,{e^{{\frac {3n_2\,{r}^{2}}{{R}^{2 }}}}} {R}^{2}{r}^{4}-{r}^{6}n_0\,n_2\,{ e^{{\frac {3n_2\,{r}^{2}}{{R}^{2}}}}} +{r}^{6}{n_0}^{2} { e^{{\frac {3n_2\,{r}^{2}}{{R}^{2}}}}}+2\,n_0\, { e^{{\frac {3n_2\,{r}^{2}}{{R}^{2}}}}}{R}^{2}{r}^{4}+192\,\alpha_2\,{R}^{6}{n_0}^{2}{e^{{\frac {n_2\,{r}^{2}}{{R}^{2}}}}}\right. \nonumber\\
&\left.+32\,\alpha_2\,{R}^{6}{n_0}^{2} { e^{{\frac {2n_2\,{r}^{2}}{{R}^{2}}}}}-2560\,{r}^{6}\alpha_2\,{n_0}^{3 }{n_2}^{2}+5376\,{r}^{6}\alpha_2\,{n_0}^{2}{n_2}^{3 }+256\,{r}^{6}\alpha_2\,{n_0}^{4}n_2+32\,\alpha_2\,{ n_0}^{4}{R}^{2}{r}^{4}+192\,\alpha_2\,{n_0}^{3}{R}^{4}{r }^{2}\right. \nonumber\\
&\left.-224\,\alpha_2\,{R}^{6}{n_0}^{2}+896\,\alpha_2\,{n_0}^{2}n_2\,{e^{{\frac {n_2\,{r}^{2}}{{R}^{2}}}}} {R}^{4}{r}^{2}-64\,\alpha_2\,{n_0}^{2}n_2\, { e^{{\frac {2n_2\,{r}^{2}}{{R}^{2}}}}}{R}^{4}{r} ^{2}+704\,\alpha_2\,{n_0}^{2}{n_2}^{2}{e^{{\frac { n_2\,{r}^{2}}{{R}^{2}}}}}{R}^{2}{r}^{4}+32\,\alpha_2\,{n_0}^{2}{n_2}^{2} { e^{{\frac {2n_2\,{r}^{2}}{{R}^{2}}}}}{R}^{2}{r}^{4}\right. \nonumber\\
&\left.-64\,\alpha_2\,{n_0}^{3}n_2\, { e^{{\frac {2n_2\,{r}^{2}}{{R}^{2}}}}}{R}^{2}{r}^{4}-128\,\alpha_2\,{n_0}^{3}n_2 \,{e^{{\frac {n_2\,{r}^{2}}{{R}^{2}}}}}{R}^{2}{r}^{4}-640\, {r}^{6}\alpha_2\,{n_0}^{2}{n_2}^{3}{e^{{\frac {n_2\,{r}^{2}}{{R}^{2}}}}}+768\,{r}^{6}\alpha_2\,{n_0}^{3} {n_2}^{2}{e^{{\frac {n_2\,{r}^{2}}{{R}^{2}}}}}\right. \nonumber\\
&\left.-128\,{r }^{6}\alpha_2\,{n_0}^{4}n_2\,{e^{{\frac {n_2 \,{r}^{2}}{{R}^{2}}}}}-64\,\alpha_2\,{n_0}^{4}{e^{{ \frac {n_2\,{r}^{2}}{{R}^{2}}}}}{R}^{2}{r}^{4}+32\,\alpha_2\, {n_0}^{4} { e^{{\frac {2n_2\,{r}^{2}}{{R}^{2}}}}}{R}^{2}{r}^{4}+64\,\alpha_2\,{n_0}^{3} { e^{{\frac {2n_2\,{r}^{2}}{{R}^{2}}}}}{R}^{4}{r} ^{2}-256\,\alpha_2\,{n_0}^{3}{e^{{\frac {n_2\,{r}^ {2}}{{R}^{2}}}}}{R}^{4}{r}^{2}\right. \nonumber\\
&\left.-1088\,\alpha_2\,{n_0}^{2}n_2\,{R}^{4}{r}^{2}-3040\,\alpha_2\,{n_0}^{2}{n_2}^{2}{R }^{2}{r}^{4}+576\,\alpha_2\,{n_0}^{3}n_2\,{R}^{2}{r}^{4} \right).
\end{align}
Furthermore, we present the force with anisotropic,  $F_a=\frac{2\Delta}{r}$, which arises from the pressure difference. In situations with significant anisotropy $0<r\leq R_s$, where $\Delta>0$, it necessitates that $p_t>p_r$ throughout the star's interior. Conversely, in the mild anisotropy case, where $\Delta<0$, it requires that $p_r>p_t$   within the pulsar.
\subsection{Junction conditions}\label{Sec:Match}

To address the equation of motions under specified junction conditions at $r = R_s$, where   $p_r = 0$, the inner metric described by Eq. (\ref{RG}) necessitates the application of these matching conditions. This can be achieved by smoothly connecting the interior metric at $r = R_s$ to Schwarzschild's exterior metric, which is expressed as:\footnote{{ To justify the use of Schwarzschild solution as an exterior solution for our system we will apply the form of Schwarzschild spacetime to the field equations of $f(G)=\alpha_1 G^2$. For this purpose, we get the following non-vanishing components:
\begin{align}
&tt-component\equiv\frac{1152\alpha_1 M^3G^3(67GM-32Rc^2)}{c^8R^6}\,, \qquad rr-component\equiv \frac{1152\alpha_1 M^3G^3(11GM-4Rc^2)}{c^8R^6}\,,\nonumber\\
 &\theta\, \theta-component\equiv\frac{1152\alpha_1 M^3G^3(41GM-18Rc^2)}{c^8R^6}\,.
\end{align}
If we calculate the above quantities numerically using the data of the pulsar under consideration which are $M=2.07 \pm 0.11 M_\odot$ and $R_s=12.34^{+1.89}_{-1.67}$ km and $\alpha_1=0.0004$  the following numerical values are obtained:
\begin{align}
&tt-component\equiv\, 7.089001308\times10^{-10}\,, \quad rr-component\,\equiv 5.866809379\times 10^{-11}\,,\quad \theta\, \theta-component\, 3.609693140\times 10^{-10}\nonumber\\
&\mbox{while  the Schwarzchild mass term has an oder} \sim 0.5\,.
\end{align}
The above numerical values indicate that we can use the Schwarzschild solution as an exterior solution for the theory under consideration.}}
\begin{equation}
 {   ds^2=-\left(1-\frac{2GM}{c^2r}\right) c^2 dt^2+\frac{dr^2}{\left(1-\frac{2GM}{c^2 r}\right)}+r^2 (d\theta^2+\sin^2 \theta d\phi^2)}\,,
\end{equation}
where, as seen from a point at an infinite distance, $M$ denotes the star's gravitational mass. We apply the junction conditions as follows:
\begin{equation}\label{eq:bo}
 \psi(r={R_s})=\ln(1-C)\,, \qquad \lambda(r={R_s})=-\ln(1-C)\,, \qquad \text{and} \, \quad {{p}_r}(r={R_s})=0\,,
\end{equation}
with $C$ being  figured as:
\begin{align}\label{comp11}
 {  C=\frac{2GM}{c^2 {R_s}}}\,,
\end{align}
With the radial pressure from Eq. \eqref{sol} and the KB ansatz \eqref{eq:KB},  the parameters { $\{n_0, n_1, n_2\}$} can be expressed with respect to $\alpha_2$  $C$s.  
\section{Constraints on astrophysics and stability derived from pulsar ${\mathcal J0740+6620}$ observations}\label{Sec:Stability}

Now, we specifically employ restrictions on $M$ and $R$ from the pulsar ${\mathcal J0740+6620}$ to calculate the quadratic gravity correction parameter's value $\alpha_2$. Additionally, to calculate the quadratic gravity correction parameter's value.  As previously stated in the introduction, it is crucial to emphasize that precise observational data plays an immensely significant role in restricting $\alpha_2$.  Initially, we need to address our selection of  ${\mathcal PSR J0740+6620 }$ to constrain the form of $f({\mathcal{G}})$ theory under consideration.

\subsection{Constraints on mass and radius derived from  ${\mathcal J0740+6620}$}\label{Sec:obs_const}

Next, we employ the  values of $M$ and $R$ of ${\mathcal PSR J0740+6620}$, as  $M=2.07 \pm 0.11 M_\odot$ and   $R_s=12.34^{+1.89}_{-1.67}$ km, as documented in \citep{Legred:2021hdx}. These measurements combine data from NICER and XMM to limit $\alpha_2$.
\begin{figure*}
\centering
\subfigure[~Mass]{\label{Fig:Mass}\includegraphics[scale=0.45]{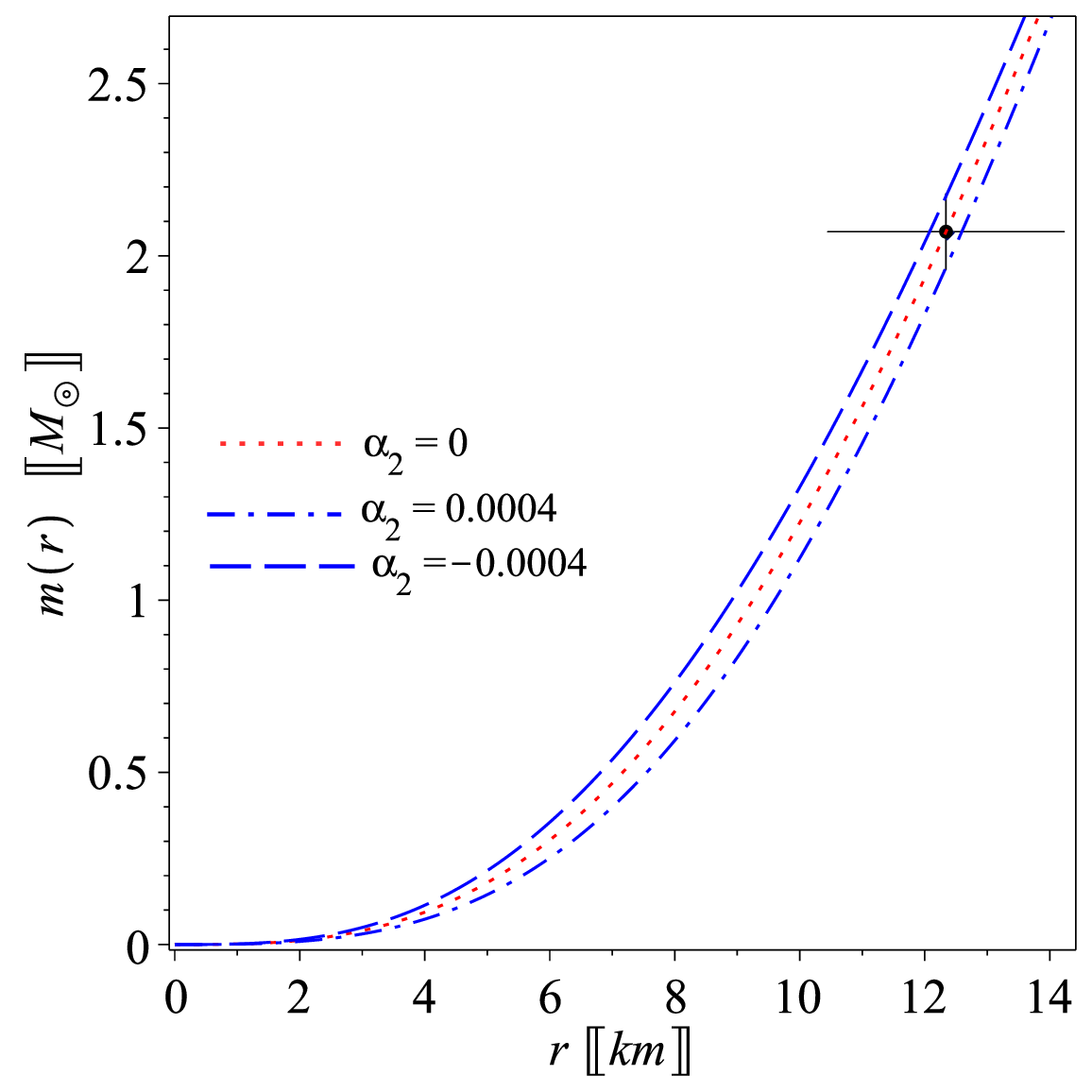}}
\caption{Mass gievn by  Eq.~\eqref{Mf3} for pulsar ${\mathcal J0740+6620}$ is shown to represent the limitations of observation on its  radius and mass ( $R_s=12.35\pm0.11$ km, $M=2.07\pm 0.11 M_\odot$) as per \citep{Legred:2021hdx}. For $\alpha_2=-0.0004$, we employ model parameters of approximately [$C\approx 0.524$, $n_0 \approx0.489$, $n_1\approx-1.174$, $n_2 =0.684$]. For $\alpha_2=0.0004$, we utilize [$C=0.47$, $n_0 =0.493$, $n_1 =-1.177$, $n_2 =0.684$]. For $\alpha_2=0$ GR, we use [$n_0 \approx 0.491$, $n_1 =-1.1756$, $n_2 =0.684$].
}
\label{Fig:Mass1}
\end{figure*}
The  mass function is defined as follows:
\begin{align}\label{Mf3}
{m(r)} =  4\pi\int_{0}^{r} \rho(\zeta) \zeta^2 d\zeta \,.
 \end{align}
Returning to the density profile described in Eq.~ \eqref{sol}, in the context of the model $R+f({\mathcal{G}})=R+\alpha_1 {\mathcal{G}}^2$, we generate the plots shown in Fig. \ref{Fig:Mass1} for various numerical value of the parameter $\alpha_2$.
\begin{itemize}
    \item In the case of $\alpha_2=0$, which corresponds to the GR  scenario, the constant's numerical values are set as follows: \{$n_0 \approx 0.491$, $n_1 =-1.1756$, $n_2 =0.684$\}.
    \item With $\alpha_2=0.0004$, we find $1.96 M_\odot$ with ${R_s}\approx12.65$ km, and  $C\approx0.47$. This leads to: \{$\alpha_2=0.0004$, $C=0.47$, $n_0 =0.493$, $n_1 =-1.177$, $n_2 =0.684$\}.
    \item When $\alpha_2=-0.0004$, the gravitational mass is approximately $2.22 M_\odot$ with ${R_s}\approx11.96$ km, and  $C\approx0.524$. These values determine the numerical constants as follows: \{$\alpha_2=-0.0004$, $C\approx 0.524$, $n_0 \approx0.489$, $n_1\approx-1.174$, $n_2 =0.684$\}.
\end{itemize}
\subsection{Geometric component }\label{Sec:geom}
\begin{figure}
\subfigure[~matching conditions]{\label{fig:Junction}\includegraphics[scale=0.38]{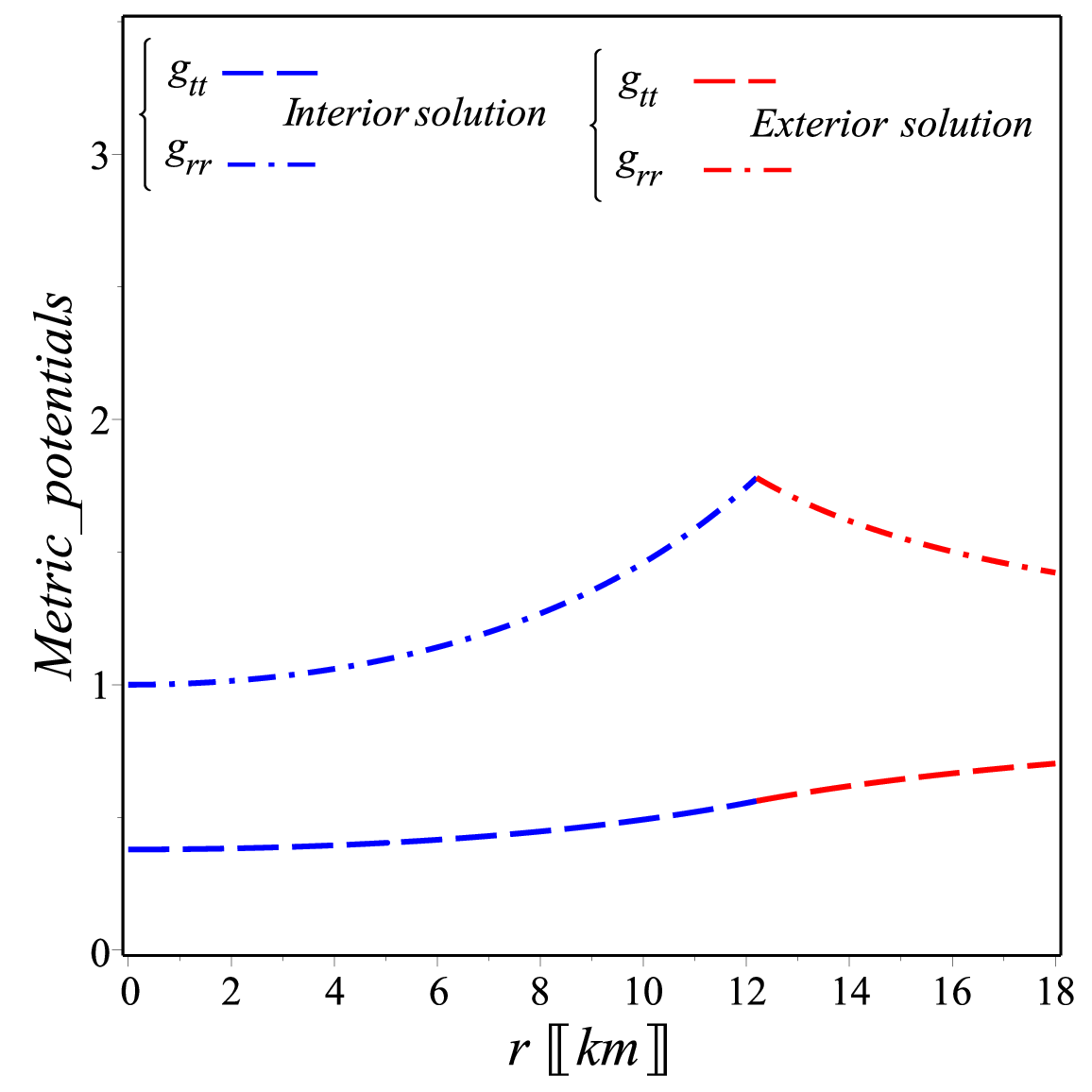}}\hspace{1cm}
\subfigure[~Z]{\label{fig:redshift}\includegraphics[scale=0.38]{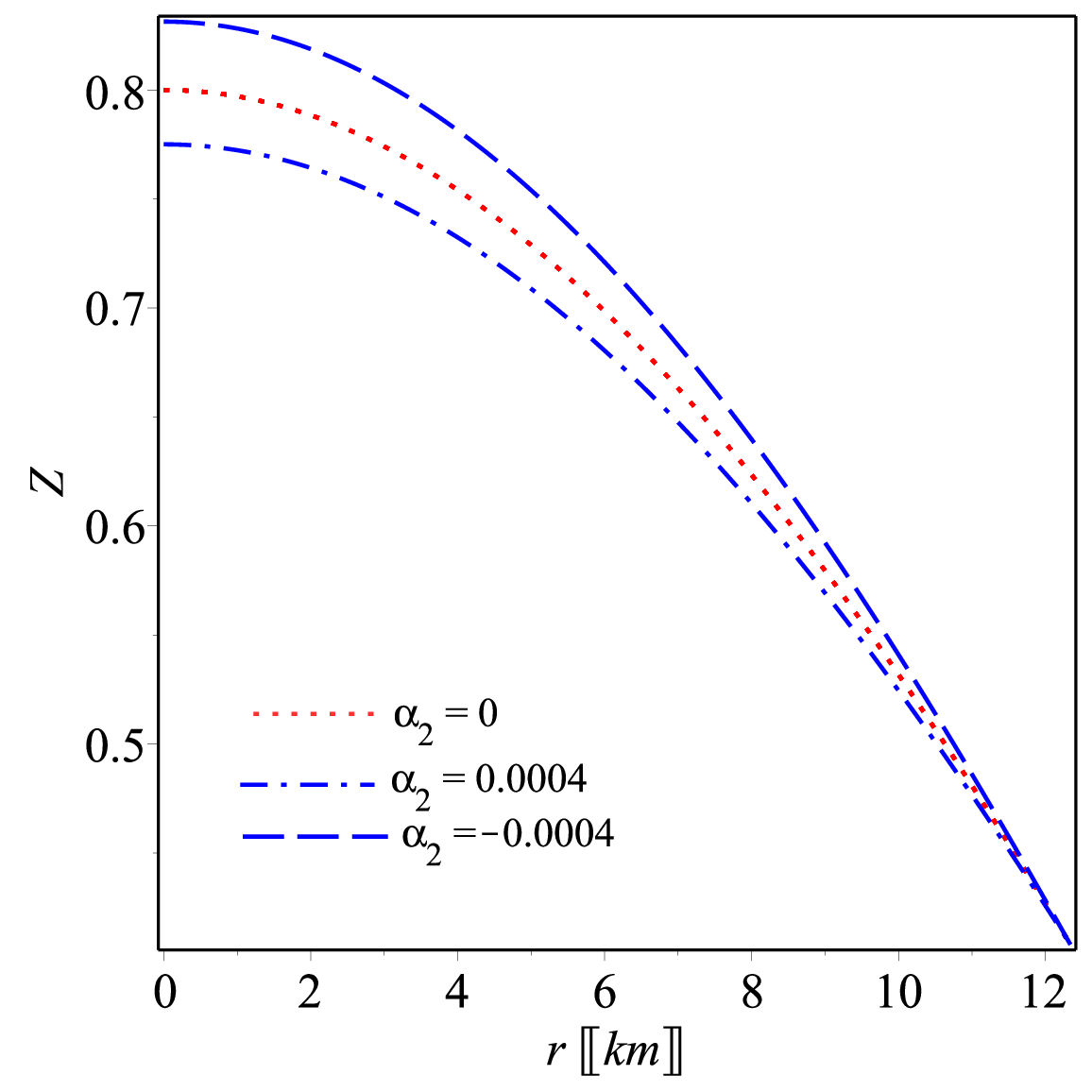}}
\caption{The geometric aspect of pulsar ${\mathcal J0740+6620}$, as depicted in \subref{fig:Junction}, illustrates $g_{tt}$ and $g_{rr}$  using the KB, and  the vacuum is the Schwarzschild exterior solution. These graphical representations serve to confirm the metric potentials' finite values throughout the interior of the pulsar and their seamless connection; \subref{fig:redshift} the redshift function \eqref{eq:redshift}, when considering values of $\alpha_2$ equal to 0, as well as approximately $\pm 0.0004$, exhibits a peak redshift at the central point $\approx 0.8$, which then consistently reduces to approximately $0.4$  as $r\to R_s$, across all scenarios.}
\label{Fig:Matching}
\end{figure}

We must highlight that the metric components ${\textit g_{tt}}$ and ${\textit g_{rr}}$ should not display any singularities within the star's interior.  The   ansatz given by Eq.~\eqref{eq:KB} guarantees  regularity of these potentials at the central region because ${ {g_{tt}(r=0)=e^{\psi}}}\neq 0$ and ${ {g_{rr}(r=0)=1}}$. Meanwhile, the way these potentials behave inside the star at any given radial position is shown in Fig.~\ref{Fig:Matching} \subref{fig:Junction}.

Additionally, we express the redshift function on the KB potentials in relation to the quadratic form of ${\mathcal{R}}+f({\mathcal{G}})$ gravity.
\begin{equation}\label{eq:redshift}
    Z(r)=\frac{1}{\sqrt{-g_{tt}}}-1=\frac{1}{\sqrt{e^{n_0 (r/R_s)^2+n_1}}}-1.
\end{equation}
We have generated $Z$ plot for  ${\mathcal J0740+6620}$, varying  $\alpha_2$ as depicted in plot \ref{Fig:Matching}\subref{fig:redshift}. When $\alpha_1=0$ (which corresponding to { GR}), $Z(r\to 0)$ yields $Z(0)\approx 0.806$, while  $Z(r\to R_s)$ is $Z_s=Z_{R_s}\approx 0.4$. When considering $\alpha_2=-0.0004$, $Z$ reaches its maximum value at the core with $Z(0)\approx 0.832$, which is greater  from GR. It then gradually reduces as we approach the surface, with $Z_s\approx 0.405$, a value close to that of  GR  which is below $2$ as indicated in \citep{Buchdahl:1959zz,Ivanov:2002xf,Barraco:2003jq,Boehmer:2006ye}. Likewise, for $\alpha_2=0.0004$, we observe the maxima of $Z(r\to 0)$ is $Z(0)\approx 0.774$, which is lower than the corresponding { GR} value. This maximum redshift then gradually diminishes as we approach the surface, with the redshift at the surface measuring approximately $Z_s\approx 0.4072$ a value close to the GR prediction.

\subsection{Matter sector}\label{Sec:matt}

\begin{figure*}
\subfigure[~$\rho$]{\label{fig:density}\includegraphics[scale=0.3]{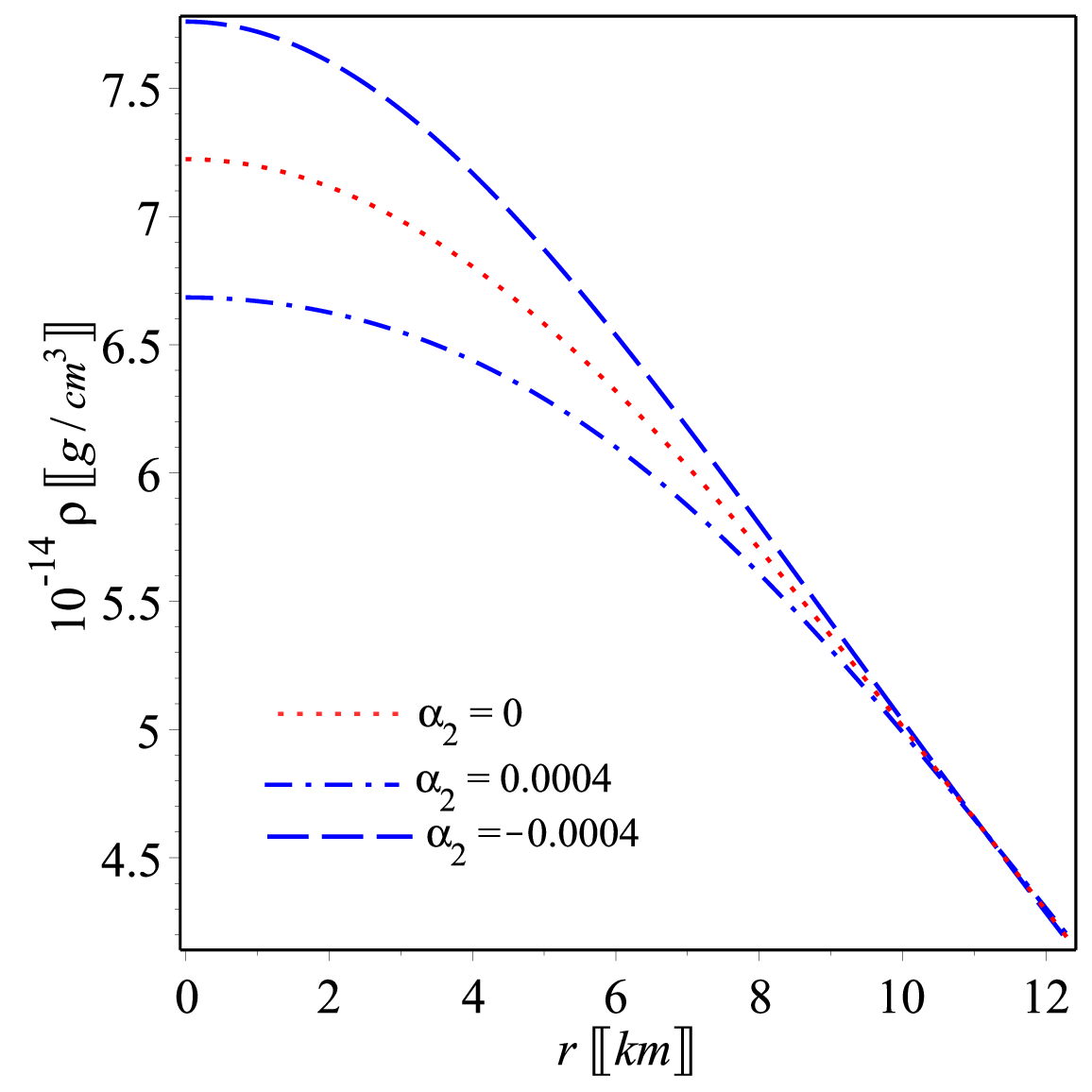}}\hspace{0.5cm}
\subfigure[~$p_r$]{\label{fig:radpressure}\includegraphics[scale=0.3]{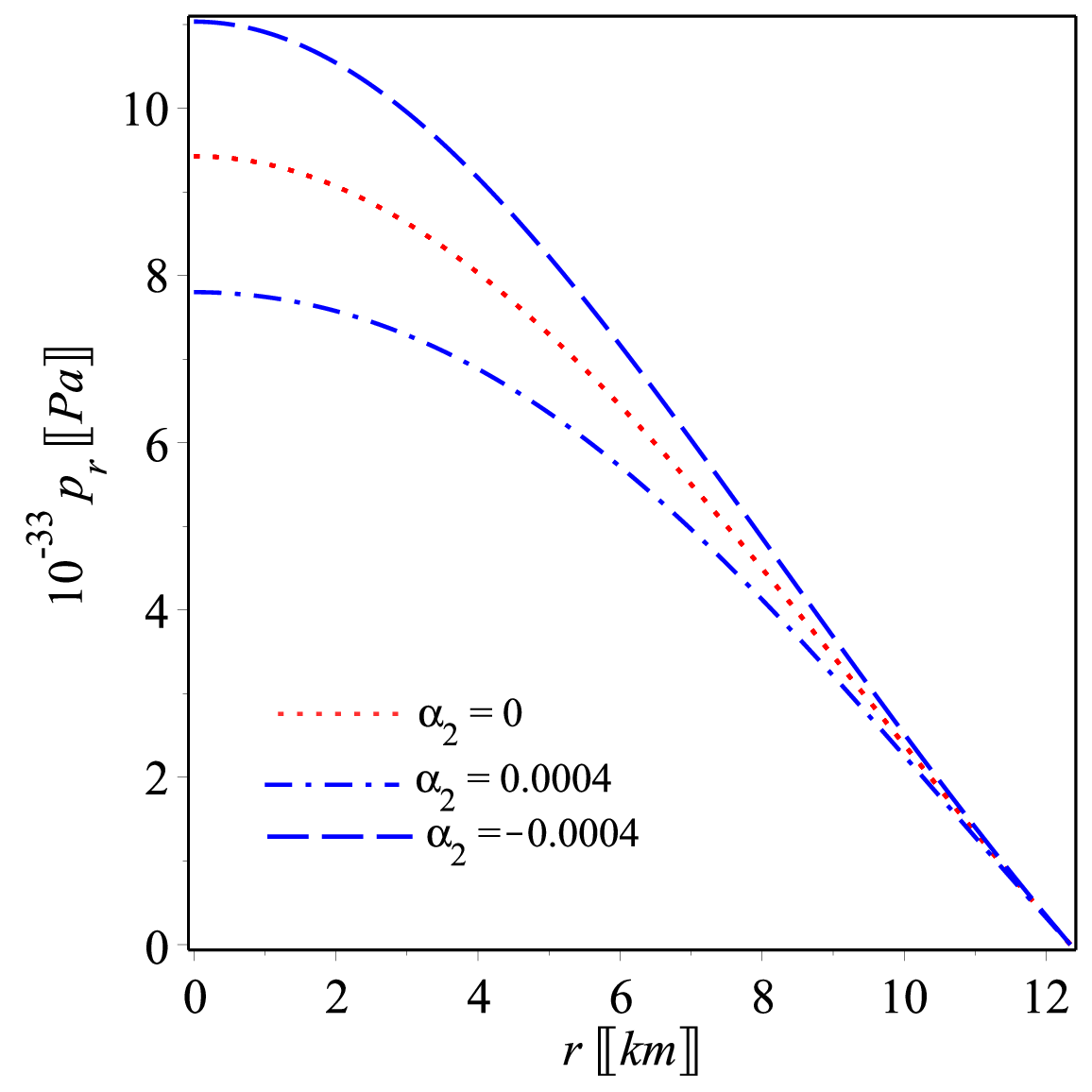}}\\
\subfigure[~ $p_t$]{\label{fig:tangpressure}\includegraphics[scale=0.3]{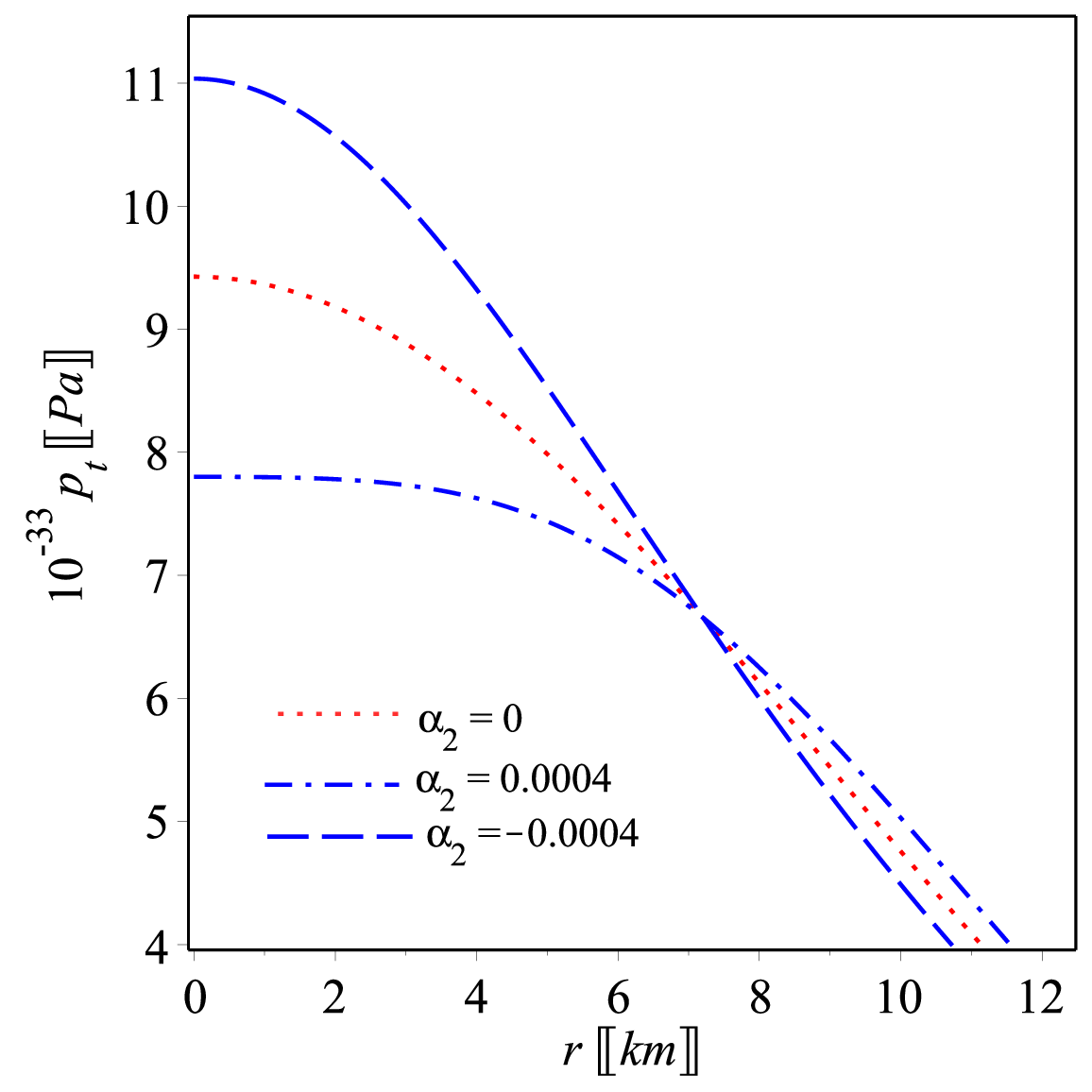}}\hspace{0.5cm}
\subfigure[~$\Delta$]{\label{fig:anisotf}\includegraphics[scale=0.3]{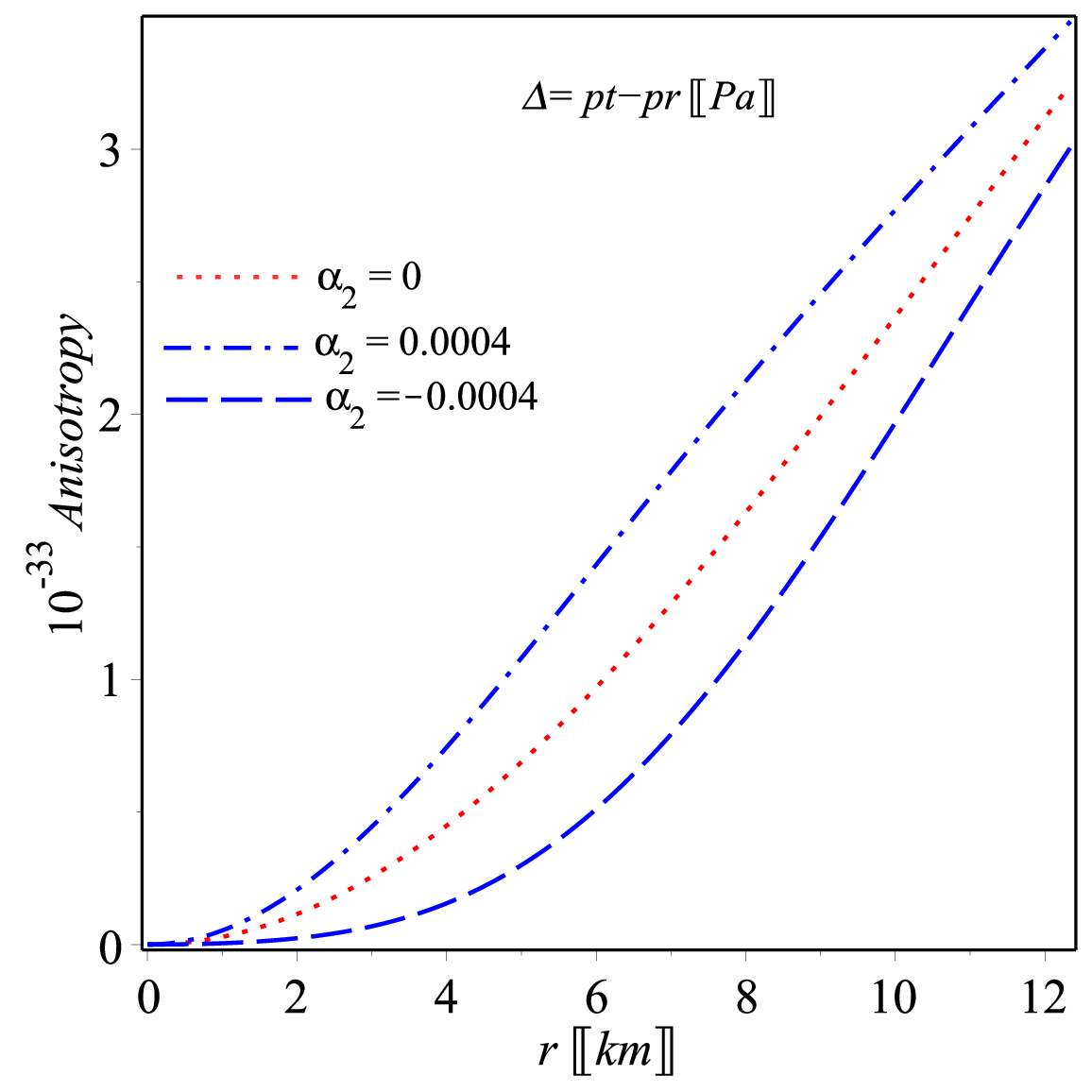}}
\caption{Characteristics of pulsar under consideration, as depicted in \subref{fig:density} to \subref{fig:tangpressure}, illustrate $\rho$, $p_r$ and $p_t$ of Eq.~\eqref{sol}. These visual representations are provided for three scenarios: $\alpha_2=0,~\pm 0.0004$. The illustrations confirm that $\rho$ and the components of the pressures maintain limited values through the inner of the pulsar and consistently reduce as we approach the surface \subref{fig:anisotf} depicts the anisotropy using the three cases: $\alpha_2=0,~\pm 0.0004$. It's evident that at the central point, the anisotropy vanishes, indicated by $p_t=p_r$. However, in regions away from the center, $\Delta$ takes on positive values (reflecting strong anisotropy with $p_t>p_r$), which, in turn, gives rise to a repulsive force denoted as $F_a=2\Delta/r$. This force plays a pivotal role in adjusting the pulsar's radius.}
\label{Fig:dens_press}
\end{figure*}

By referring to Equ.~\eqref{sol} in conjunction with the numerical values in Subsection \ref{Sec:obs_const}, we can generate visual representations of$\rho$, $p_r$ and $p_t$ as functions of the radial coordinate, as shown in the Fig.~\ref{Fig:dens_press}\subref{fig:density}--\subref{fig:tangpressure}. Clearly, $rho$, $p_r$ and $p_t$ adhere to the stability criteria.
 Furthermore, we graphically represent  $\Delta(r)$,  in Fig.~\ref{Fig:dens_press}\subref{fig:anisotf}.
 It's important to highlight that in situations of substantial anisotropy, such as in our current investigation, an additional positive force, proportional to $\Delta/r$, becomes a significant factor. This opposing force counteracts gravity, allowing the star to support high mass and achieve enhanced compactness which we will provide in details in Subsection \ref{Sec:TOV}.

It's important to provide numerical values for certain physical parameters of pulsar ${\mathcal J0740+6620}$, as projected by the current model. These values are as follows: With $\alpha_2=0.0004$, the core density amounts to approximately ${\rho_\text{core}\approx 6.68\times 10^{14}}$ g/cm$^{3}$, which is approximately 2.5($\rho_\text{nuc}$), and  $p_{r\to 0}$, is roughly $7.81\times 10^{33}$ dyn/cm$^2$, approximately equal to the core tangential pressure, $p_{t\text{(core)}}$. At the pulsar's surface, the density is approximately ${\rho_s\approx 4.14\times 10^{14}}$ g/cm$^{3}$, which is approximately 1.5($\rho_\text{nuc}$). The radial pressure at the surface, $p_{r(R_s)}$, is approximately 0 dyn/cm$^2$, and the tangential pressure at the surface, $p_{t(R_s)}$, is approximately $4.01\times 10^{33}$ dyn/cm$^2$.

For $\alpha_2=-0.0004$, the central density amounts to approximately ${\rho_\text{core}\approx 7.78\times 10^{14}}$ g/cm$^{3}$, which is approximately 2.9($\rho_\text{nuc}$). The pressure $p_{r \to 0}$, is approximately $1.104\times 10^{34}$ dyn/cm$^2$, approximately equal to the core tangential pressure, $p_{t\text{(core)}}$. At the pulsar's surface, we observe the density to be approximately ${\rho_s\approx 4.1\times 10^{14}}$ g/cm$^{3}$, which is roughly 1.4 times the nuclear density ($\rho_\text{nuc}$). The radial pressure at the surface, $p_{r(R_s)}$, is nearly 0 dyn/cm$^2$, while the tangential pressure at the surface, $p_{t(R_s)}$, is approximately $3.9\times 10^{33}$ dyn/cm$^2$.

As explained in Section \ref{Sec:Model}, we have employed the KB ansatz \eqref{eq:KB} to close the system \eqref{Rho1}--\eqref{pt1} instead of relying on EoS's.
Consequently, we define $\eta:=r/R_s$ and express asymptotically Eqs. \eqref{sol}  to order $O(\eta^2)$. These equations yield the following relationships:
\begin{equation}\label{eq:KB_EoS}
  \mathrm{ p_r(\rho)\approx c_1 \rho+c_2\,, \qquad  p_t(\rho) \approx c_3 \rho+c_4}\,,
\end{equation}
where $c_1$, $c_2$, $c_3$, and $c_4$, are entirely dictated by specific parameters in the model space, which are defined in Appendix  \ref{Sec:App_1}.  Notably, we can represent the previously mentioned equations in a way that is more intuitively understandable from a physical perspective.
\begin{equation}\label{eq:KB_EoS2}
  {\mathrm  p_r(\rho)\approx v_r^2(\rho-{ \rho_1})\,, \qquad  p_t(\rho) \approx v_t^2 (\rho-{ \rho_2})}\,,
\end{equation}
In this case, the quantities assume a more comprehensible form:  $v_t^2=c_1$,  $v_r^2=c_3$ are the speeds of sound in the tangential and radial directions moreover,  $c_2/c_1$ is the density defined as $-{ \rho_1}$ and so on. It's worth noting that ${ \rho_1}$ corresponds  $\rho_s$ and verifies $p_r(\rho_s)=0$.    This doesn't hold true to ${ \rho_2}$ because $p_t$  may not necessarily be zero at the surface.
\subsection{Zeldovich's criterion}
A crucial requirement to ensure the stability of a star, as discussed in \citep{1971reas.book.....Z}, is that $p_r(r\to 0)$ should not  $\rho(r\to 0)$, i.e.
\begin{equation}\label{eq:Zel}
   {\mathrm  {\frac{{p}_r(0)}{c^2{\rho}(0)}\leq 1.}}
\end{equation}
Referring to Eq.~\eqref{sol}, we can derive $\rho(r\to 0)$ and $p_r(r\to 0)$.
\begin{align}
 & {\rho}_{_{r\to 0}}=3{\frac {n_2\, \left( 1+32\alpha_2\,{n_0}^{2}n_2-3226\,\alpha_2\,{n_2}^{2}n_0 \right) }{{c}^{2}{R_s}^{2}{ \kappa^2}}}\,, \nonumber \\
 &  {p_r}_{_{r\to0}} ={\frac {2\,n_0-1376\,\alpha_2\,{n_0}^{2}{n_2}^{2}+ \left(
256\,\alpha_2\,{n_0}^{3}-1 \right) n_2}{{
\chi}^{2}{R}^{2}}}
.\quad
&\end{align}
Employing the previously obtained numerical values for the pulsar ${\mathcal J0740+6620}$ in Subsection \ref{Sec:obs_const}, when $\alpha_2=0.0004$, the Zeldovich inequality \eqref{eq:Zel} translates to $\frac{{p}_r(0)}{c^2{\rho}(0)}=0.102$, which is less than 1. Similarly, for $\alpha_2=-0.0004$, the inequality leads to $\frac{{p}_r(0)}{c^2{\rho}(0)}=0.105$, which is also less than 1.  This affirms that the Zeldovich criterion is satisfied in both scenarios.  { One of the most exciting issues is the sign of the matter trace inside neutron stars. Usually, the trace is assumed
to be negative for normal matter (electromagnetic field and non-interacting particles). However, this condition
could be broken in strongly interacting systems as mentioned by Zeldovich . This case
within the theory under consideration, $\mathcal{R}+f(\mathcal{G})$  gravity, is not satisfied, i.e., the trace of the solution presented in this study has a positive value which is in agreement with what presented in \cite{Podkowka:2018gib}.}

\subsection{Conditions related to energy distributions}\label{Sec:Energy-conditions}

It is advantageous to express the Eq.~\eqref{fe} as:
\begin{equation}\label{eq:fR_MG}
    G_{\mu\nu}=\kappa\left(\mathfrak{T}_{\mu\nu}+\mathfrak{T}_{\mu\nu}^{geom}\right)=\kappa {{T_{\mu \nu}}^{eff}}.
\end{equation}
{ Here, $G_{\mu\nu}:=\mathcal{R}_{\mu\nu}-g_{\mu\nu}\mathcal{R}/2$ represents the Einstein tensor, which accounts for the corrections introduced by the quadratic form in $f({\mathcal{G}})$ theory \cite{DeFelice:2010aj}}
\begin{align}
  &  \mathfrak{T}_{\mu\nu}^{geom}=-\frac{1}{\kappa^2}\left([2\mathcal{R}g_{\zeta\eta}\nabla^{2}+
2\mathcal{R}\nabla_{\zeta}\nabla_{\eta}+4g_{\zeta\eta}\mathcal{R}^{\mu\nu}\nabla_{\mu}\nabla_{\nu}+
4\mathcal{R}_{\zeta\eta}\nabla^{2}-4\mathcal{R}^{\mu}_{\zeta}\nabla_{\eta}\nabla_{\mu}-4\mathcal{R}^{\mu}_{\eta}
\nabla_{\zeta}\nabla_{\mu}-4\mathcal{R}_{\zeta\mu\eta\nu}\nabla^{\mu}\nabla^{\nu}]
f_{\mathcal{G}}\right.\nonumber\\
&-
\left.\frac{1}{2}g_{\zeta\eta}f+[2\mathcal{R}\mathcal{R}_{\zeta\eta}-4\mathcal{R}^{\mu}_{\zeta}\mathcal{R}_{\mu\eta}
-4\mathcal{R}_{\zeta\mu\eta\nu}R^{\mu\nu}+2\mathcal{R}^{\mu\nu\delta}_{\zeta}\mathcal{R}_{\eta\mu\nu\delta}]
f_{\mathcal{G}}\right),
\end{align}
and ${{T_{\mu \nu}}^{eff}}$ is given by ${{T_{\mu \nu}}^{eff}}=diag(-{{\color{blue} {\tilde \rho_1}}} c^2, {\tilde p_1}{}_r,  {\tilde p_1}{}_t, {\tilde p_1}{}_t)$.
\begin{figure*}
\subfigure[~WEC \&NEC (in the tangential direction)]{\label{fig:Cond2}\includegraphics[scale=0.27]{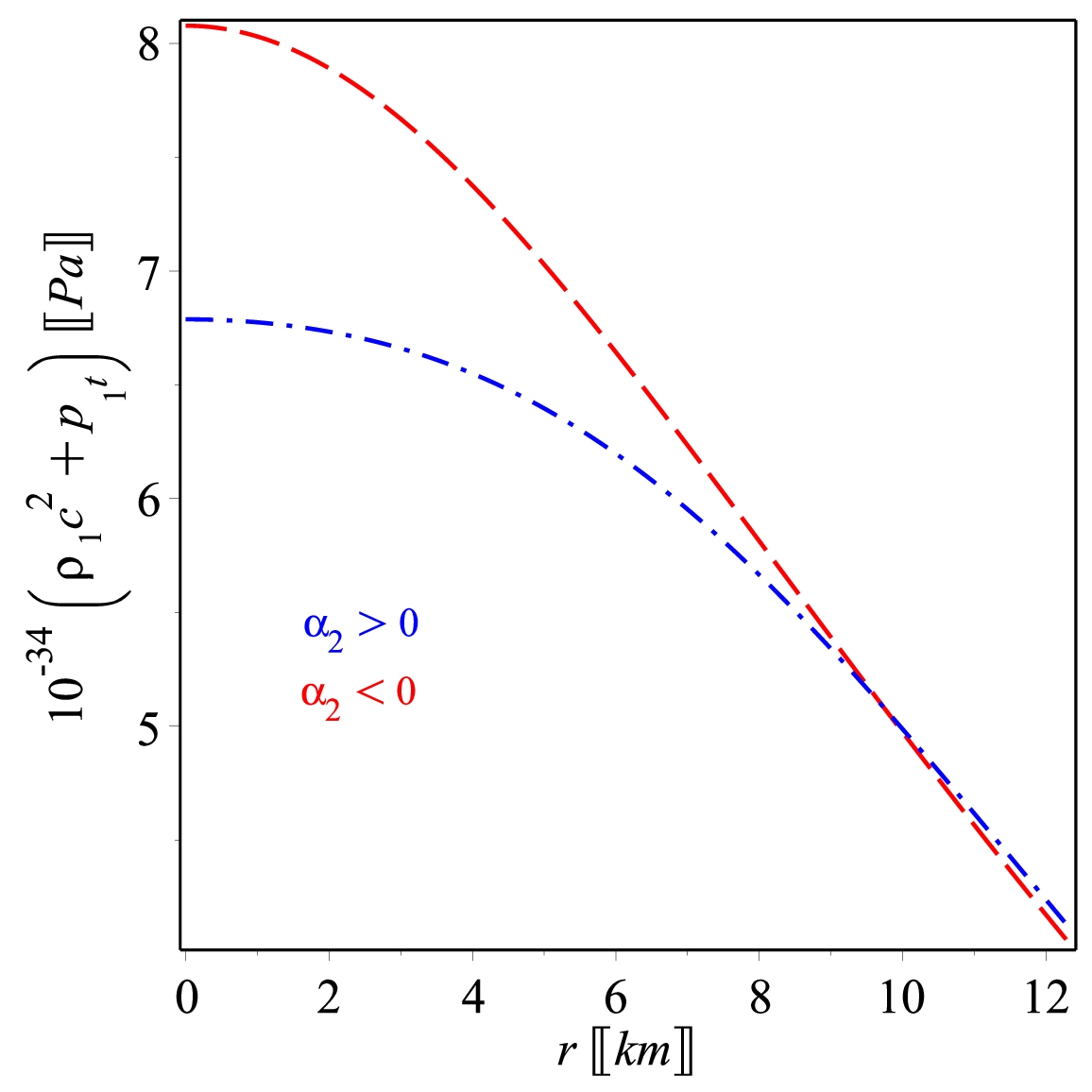}}\hspace{0.2cm}
\subfigure[~WEC \& NEC (in the radial direction)]{\label{fig:Cond1}\includegraphics[scale=0.27]{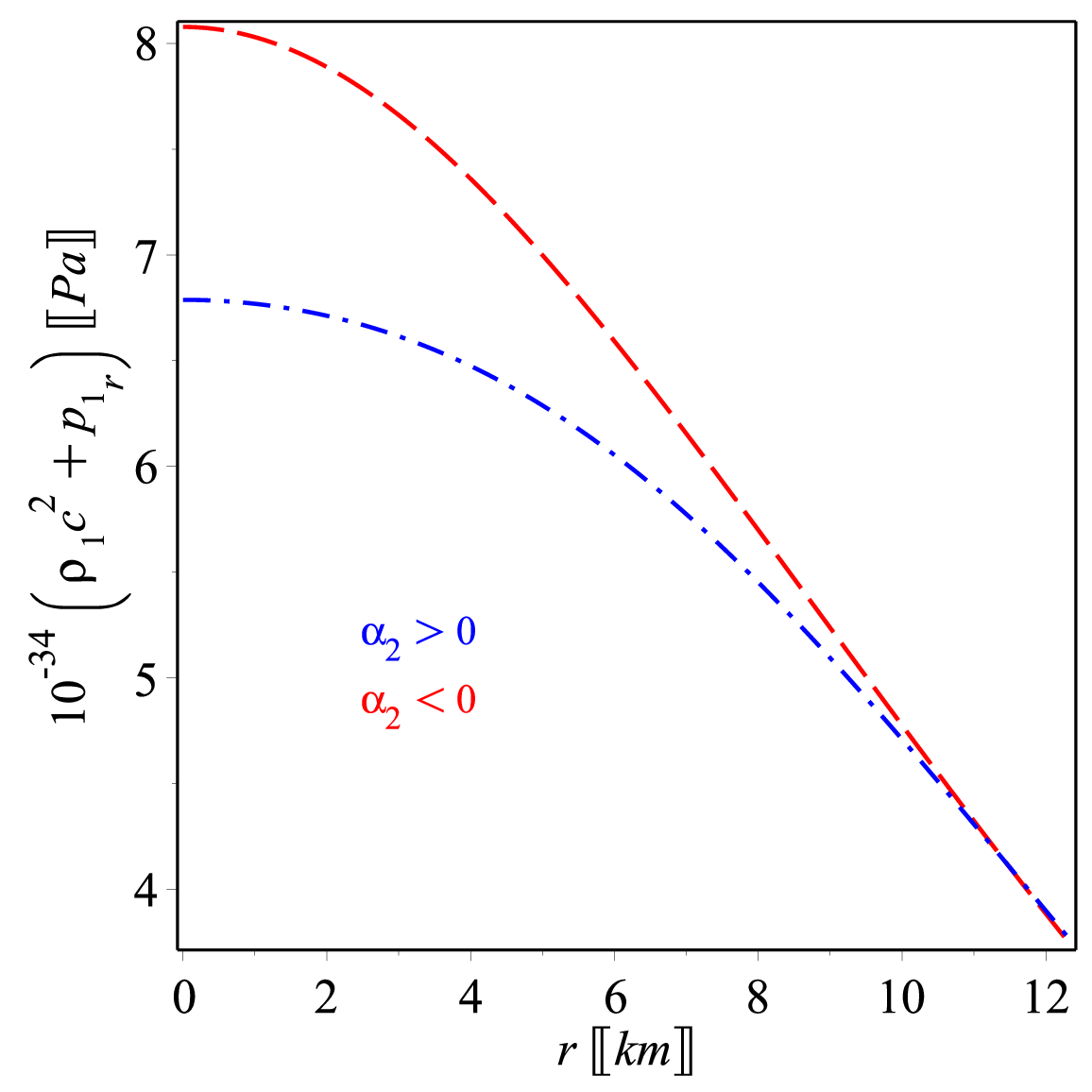}}\hspace{0.2cm}
\subfigure[~SEC]{\label{fig:Cond3}\includegraphics[scale=.27]{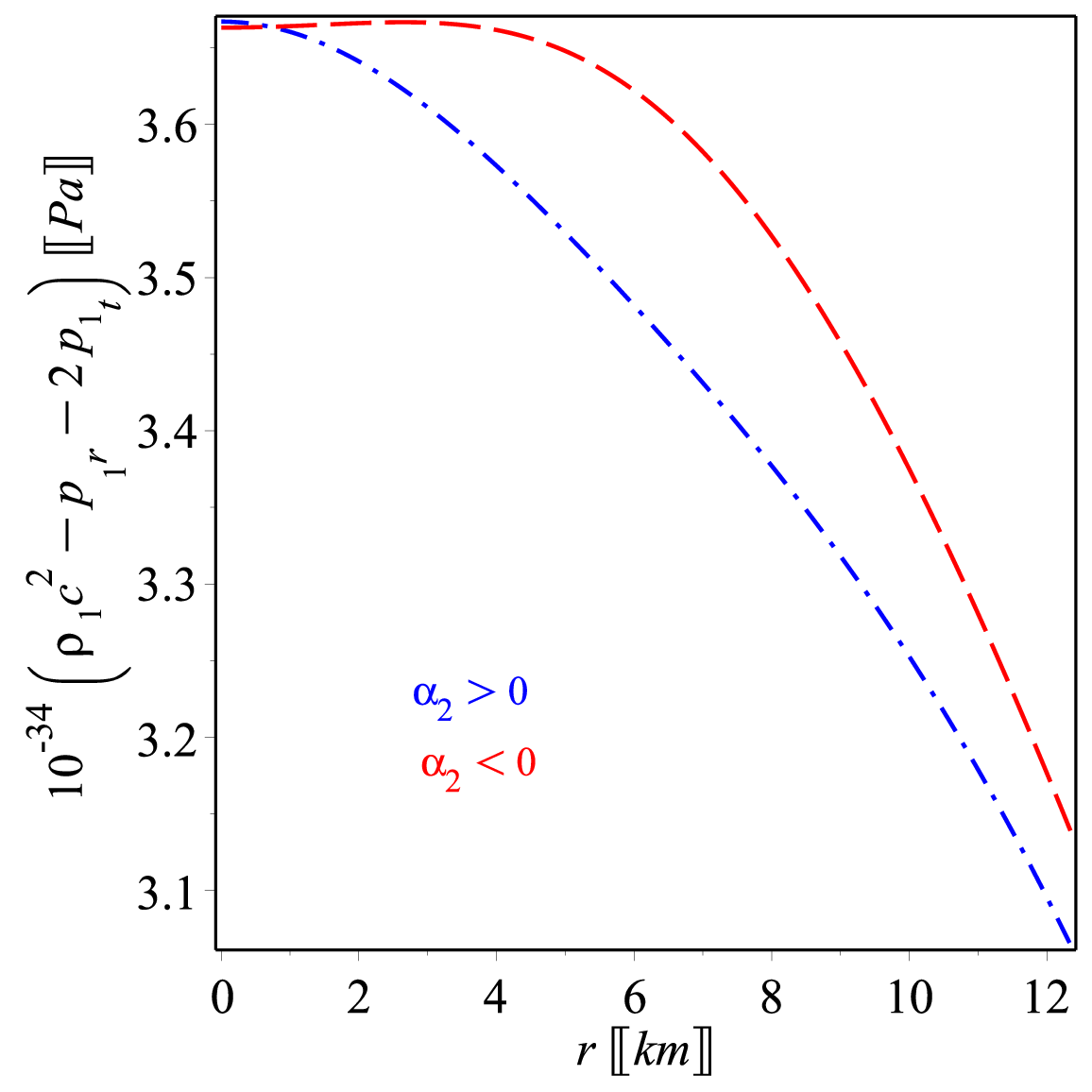}}\\
\subfigure[~DEC in the radial direction]{\label{fig:DEC}\includegraphics[scale=.27]{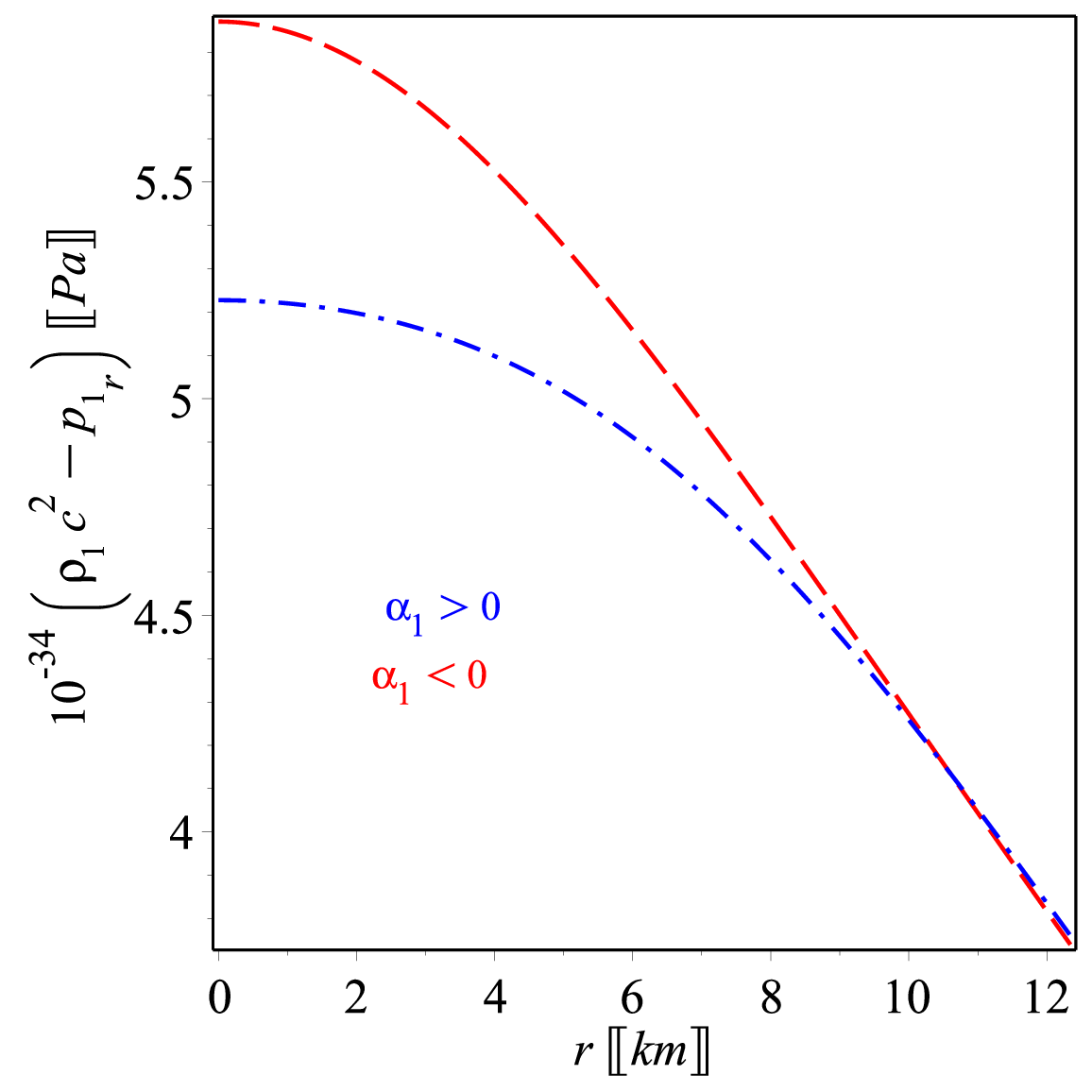}}\hspace{0.2cm}
\subfigure[~DEC in the tangential direction]{\label{fig:DEC}\includegraphics[scale=.27]{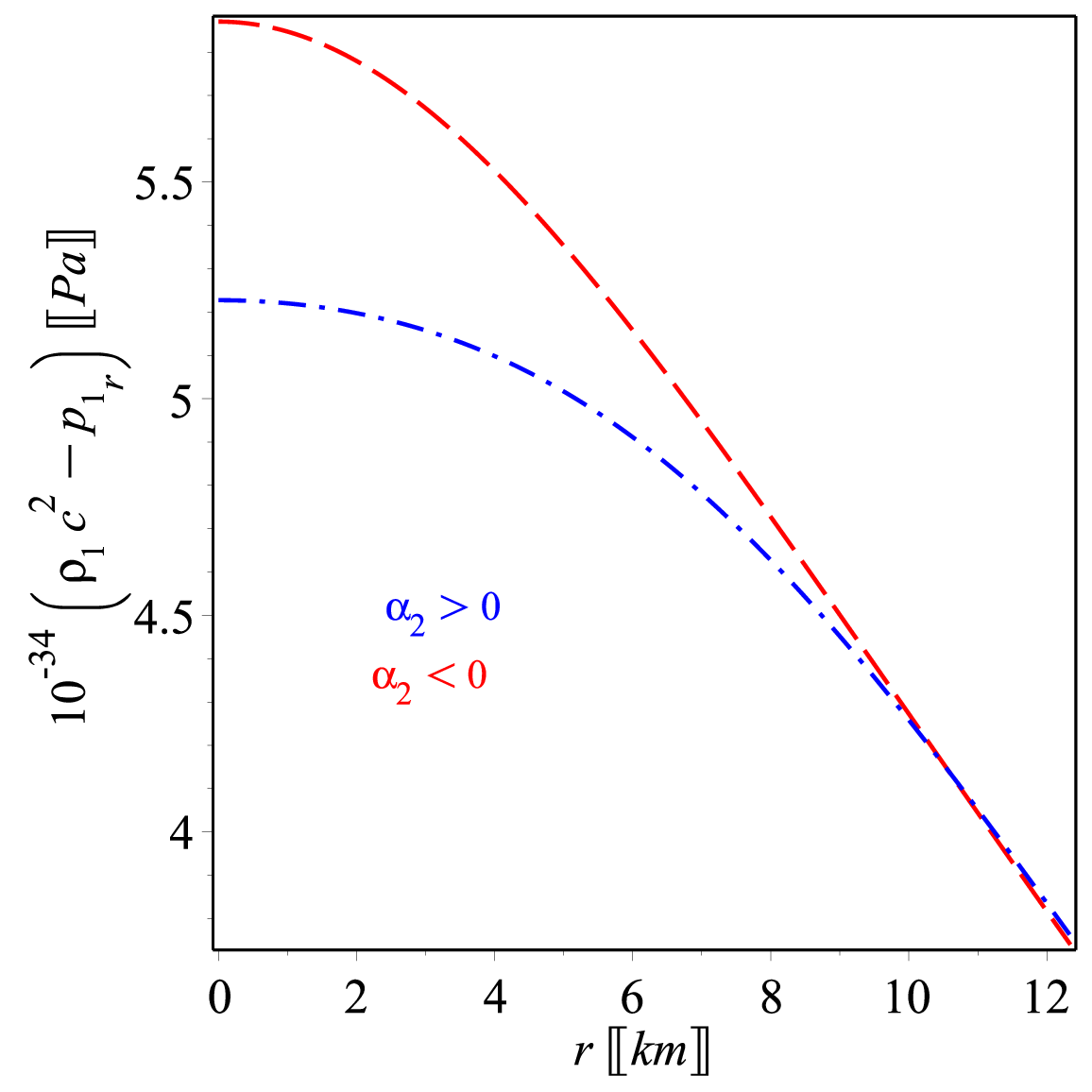}}
\caption{The graphical representations confirm the satisfaction of all the energy conditions of  ${{T_{\mu \nu}}^{eff}}$, as elaborated in Subsection \ref{Sec:Energy-conditions}. The scenarios where $\alpha_2>0$ and $\alpha_2<0$ correspond to the cases with $\alpha_2=0.0004$ and $\alpha_2=-0.0004$ respectively.}
\label{Fig:EC}
\end{figure*}
 It's important to note that the Ricci tensor in modified gravity of $\mathcal{R}+f(\mathcal{G})$ can be expressed as follows $\mathcal{R}_{\mu\nu}=\kappa\left({{T_{\mu \nu}}^{eff}}-\frac{1}{2} g_{\mu\nu} {{T_{\mu \nu}}^{eff}}\right)$. This can be recognized from  Eq.~ \eqref{eq:fR_MG}. In this context, we can modify the energy conditions to encompass  $\mathcal{R}+f(\mathcal{G})$ as:
\begin{itemize}
  \item[a.] ${{ {\tilde \rho_1}}}\geq 0$, $ {{ {\tilde \rho_1}}} c^2+ {{ {\tilde p_1}}}{}_r > 0$ and ${{ {\tilde \rho_1}}} c^2+{{ {\tilde p_1}}}{}_t > 0$, specifically, (WEC).
  \item[b.] ${{ {\tilde \rho_1}}} c^2+ {{ {\tilde p_1}}}{}_r \geq 0$ and ${{ {\tilde \rho_1}}} c^2+ {{ {\tilde p_1}}}{}_t \geq 0$, namely,   (NEC).
  \item[c.] ${{ {\tilde \rho_1}}} c^2+{{ {\tilde p_1}}}{}_r+2{{ {\tilde p_1}}}{}_t\geq 0$, ${{ {\tilde \rho_1}}} c^2+{{ {\tilde p_1}}}{}_r \geq 0$ and ${{ {\tilde \rho_1}}} c^2+{{ {\tilde p_1}}}{}_t \geq 0$.
  \item[d.] ${{ {\tilde \rho_1}}}\geq 0$, ${{ {\tilde \rho_1}}} c^2-{{ {\tilde p_1}}}{}_r \geq 0$ and ${{ {\tilde \rho_1}}} c^2-{{ {\tilde p_1}}}{}_t \geq 0$.\\
\end{itemize}
In Fig. \ref{Fig:EC}, we plot the above conditions with possible values of  $\alpha_2$. These plots confirm that the current model for pulsar ${\mathcal J0740+6620}$ is satisfied.

\subsection{Causality properties of the model}\label{Sec:causality}

\begin{figure*}
\subfigure[~Speed of sound in the radial direction]{\label{fig:vr}\includegraphics[scale=0.28]{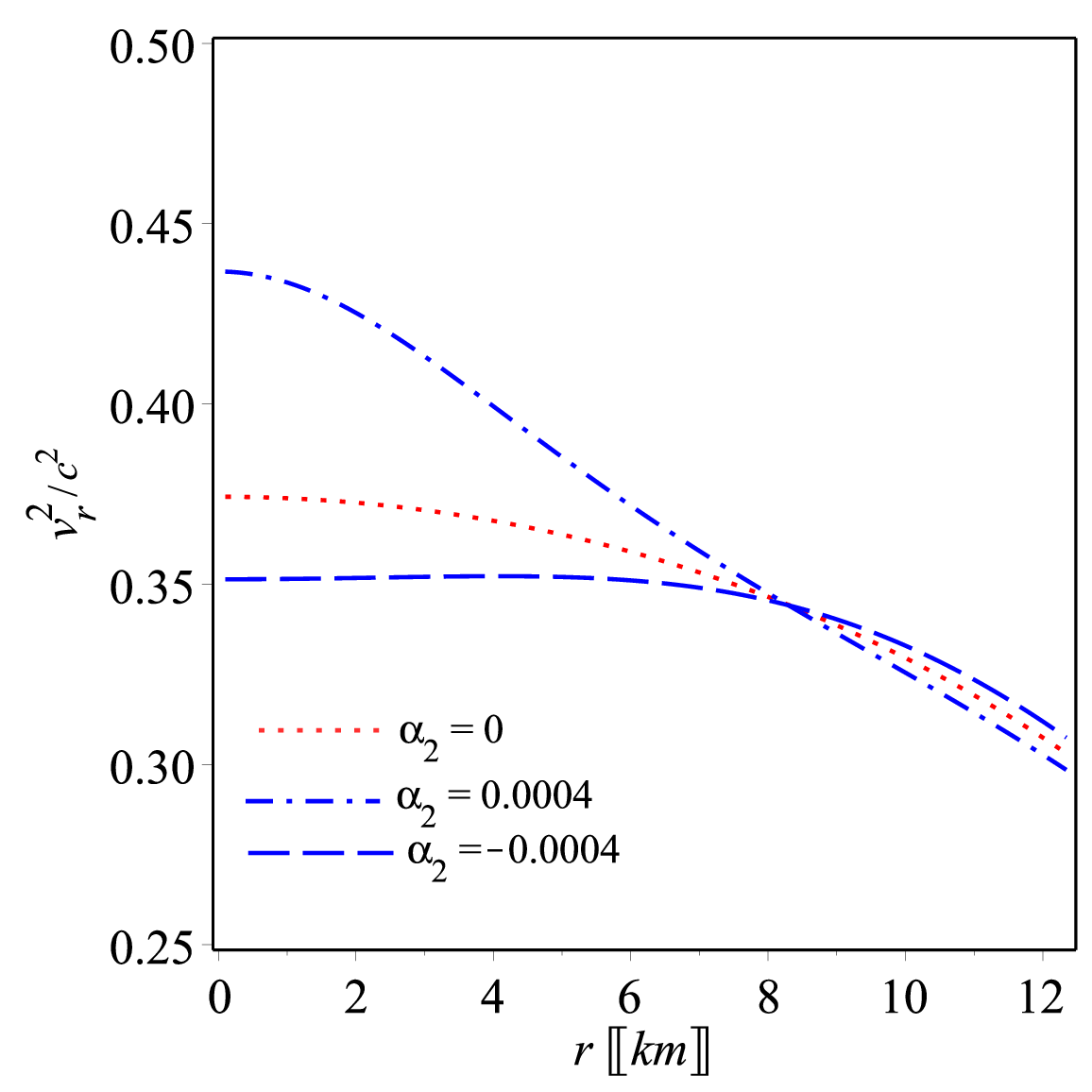}}\hspace{0.2cm}
\subfigure[~Speed of sound in the tangential direction]{\label{fig:vt}\includegraphics[scale=.28]{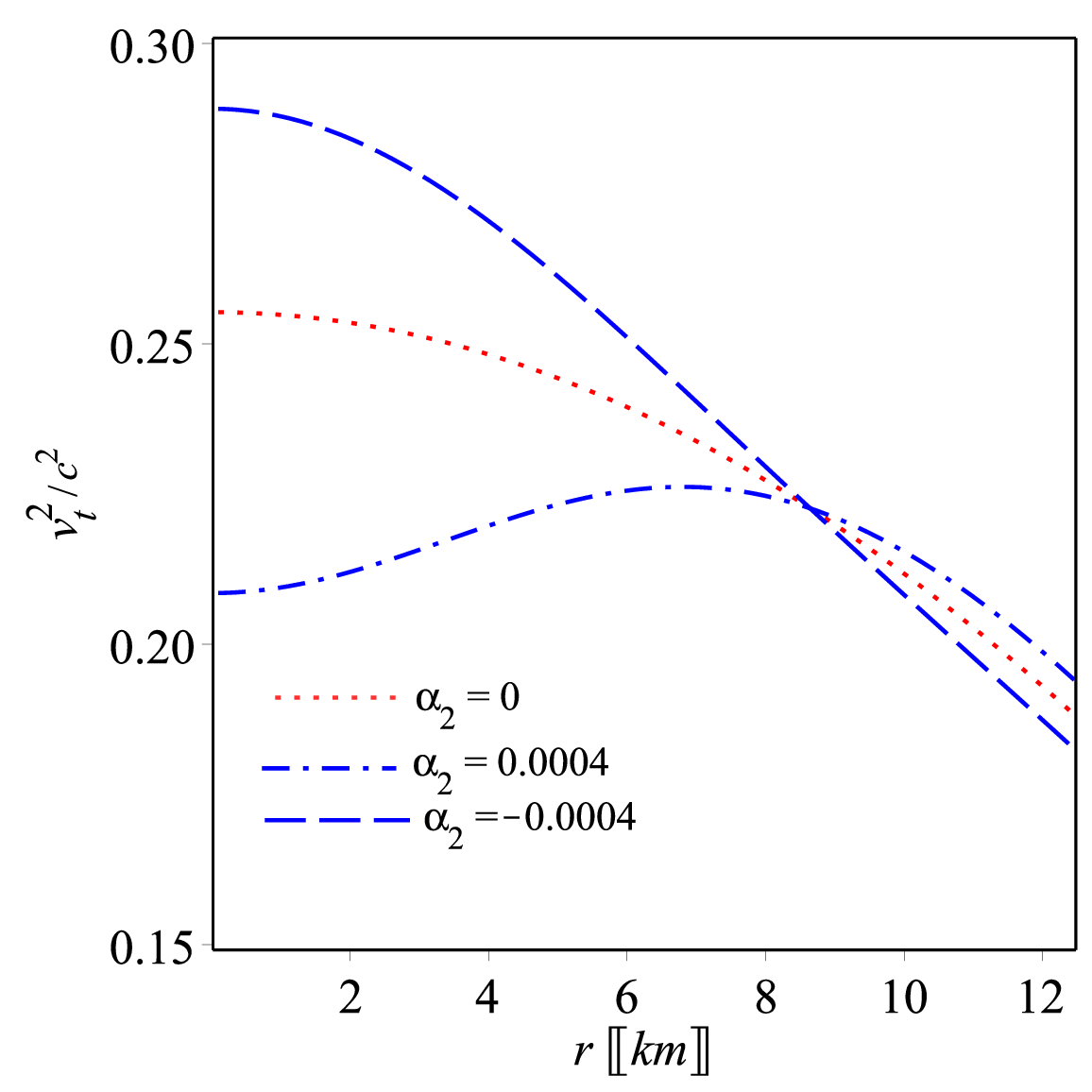}}\hspace{0.2cm}
\subfigure[~Stability under conditions of strong anisotropy]{\label{fig:vt-vr}\includegraphics[scale=.28]{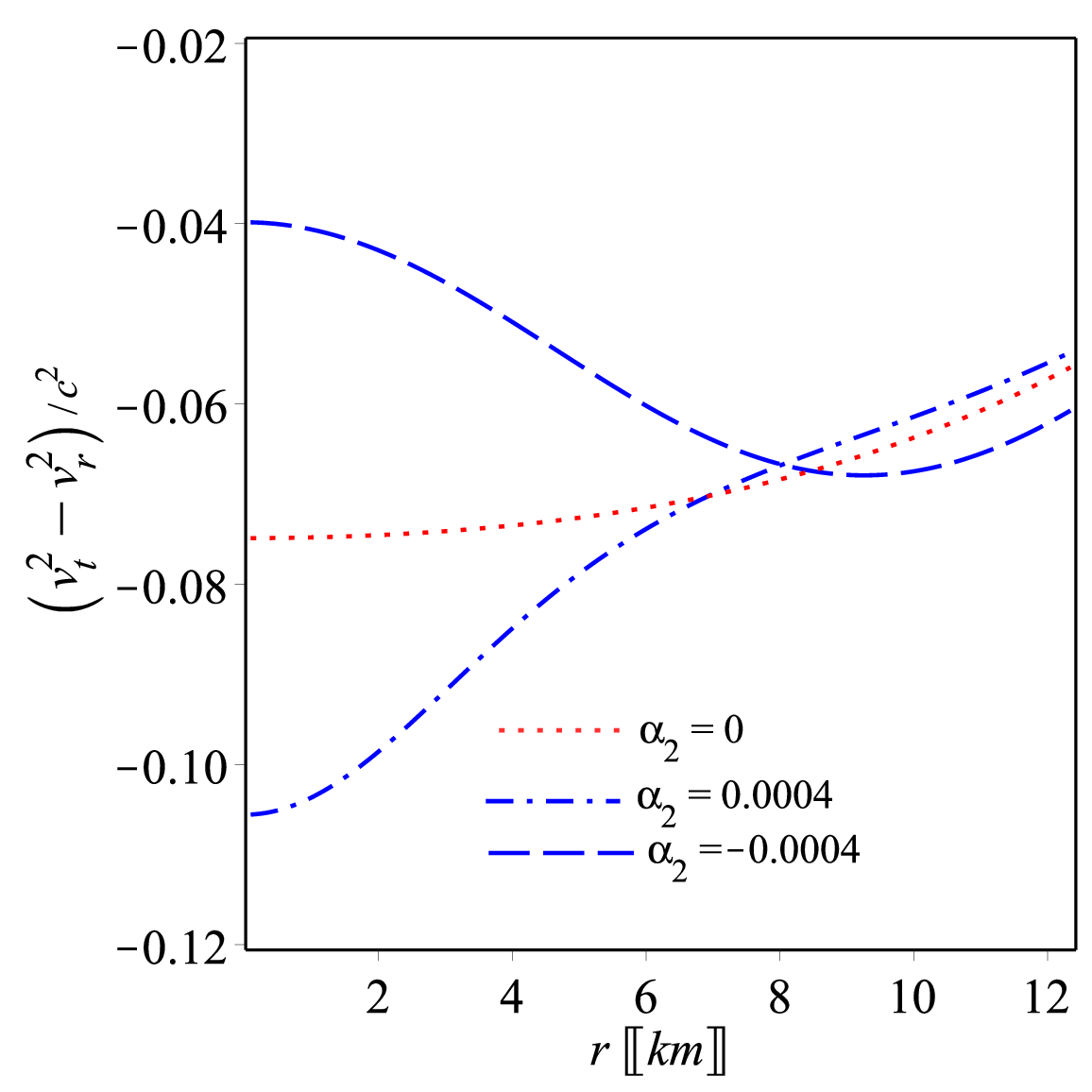}}
\caption{Sound speed   of   ${\mathcal J0740+6620}$ for $\alpha_2=0,~ \pm 0.0004$ is presented as follows: \subref{fig:vr} and \subref{fig:vt} illustrate the sound propagation  as described by Eq.~\eqref{eq:sound_speed}. In \subref{fig:vt-vr}, the figures demonstrate that $(v_t^2-v_r^2)/c^2 < 0$ required for a strongly anisotropic configuration.}
\label{Fig:Stability}
\end{figure*}

 Reflecting on the derived EoS \eqref{eq:KB_EoS2}, $p_r$ and $p_t$ are represented as:
\begin{equation}\label{eq:sound_speed}
  v_r^2 =  \frac{ d{ p}_r}{d { \rho}}=  \frac{p'_r}{{ \rho'}}, \quad
  v_t^2 = \frac{d{  p}_t}{d{   \rho}}= \frac{p'_t}{{ \rho'}}.
\end{equation}
{ Using Eq.~\eqref{sol}, we derive derivatives of $\rho$, $p_r$ and $p_t$}, which are presented in  the Appendix \ref{Sec:App_2}, as detailed in Eqs. \eqref{eq:dens_grad}--\eqref{eq:pt_grad}. We illustrate the propagation of $c_s$ in both tangential an radial directions within pulsar ${\mathcal J0740+6620}$, considering various values of the model parameter $\alpha_2$. These visual representations are provided in plots \ref{Fig:Stability}\subref{fig:vr} and \subref{fig:vt} which confirm that the values of $v_r^2/c^2$ and $v_t^2/c^2$ fall within the range of $0\leq {v_r^2}/c^2\leq 1$ and $0\leq {v_t^2/c^2} \leq 1$, thereby satisfying the causality and stability conditions.  
As Fig. \ref{Fig:Stability}\subref{fig:vt-vr} shows the stability of an anisotropic stellar configuration is verified \citep{Herrera:1992lwz}.

It's worth mentioning that the speed of sound both  directions, (radial and tangential), undergoes changes as we move radially, as visualized in plots \ref{Fig:Stability}\subref{fig:vr} and \subref{fig:vt}. When $\alpha_2=0.0004$, we observe that $0.30< v_r^2/c^2<0.45$, and $0.19<v_t^2/c^2<0.22$. In the case of $\alpha_2=-0.0004$, we find that $0.33< v_r^2/c^2<0.35$ and $0.19<v_t^2/c^2<0.30$. The maximum bounds of these periods correspond to the speeds of sound as $r\to 0$.


\subsection{Adiabatic indices and the state of hydrostatic equilibrium}\label{Sec:TOV}

\begin{figure}
\subfigure[~$\gamma$]{\label{fig:gamar1}\includegraphics[scale=0.28]{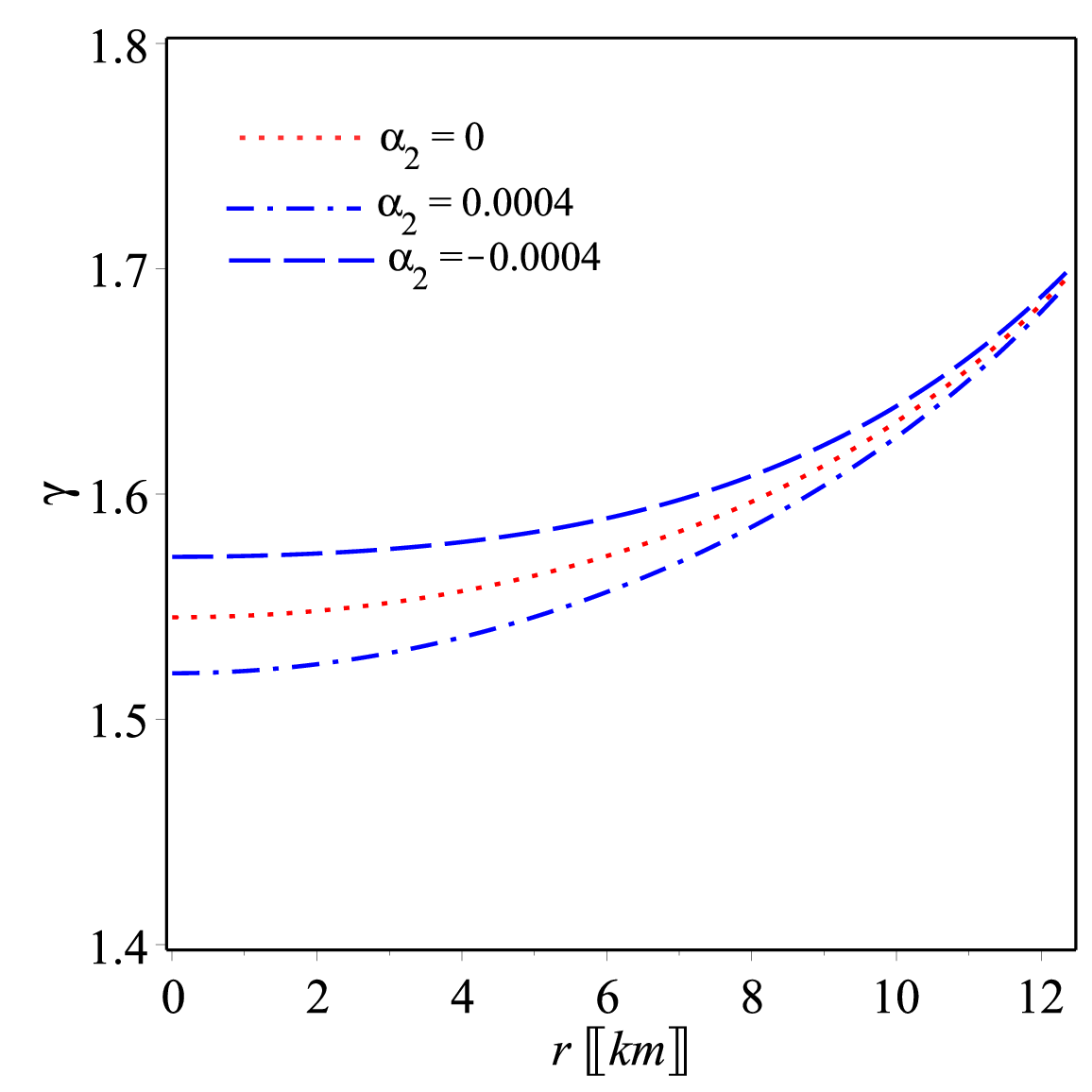}}\hspace{0.2cm}
\subfigure[~$\Gamma_r$]{\label{fig:gamar}\includegraphics[scale=0.28]{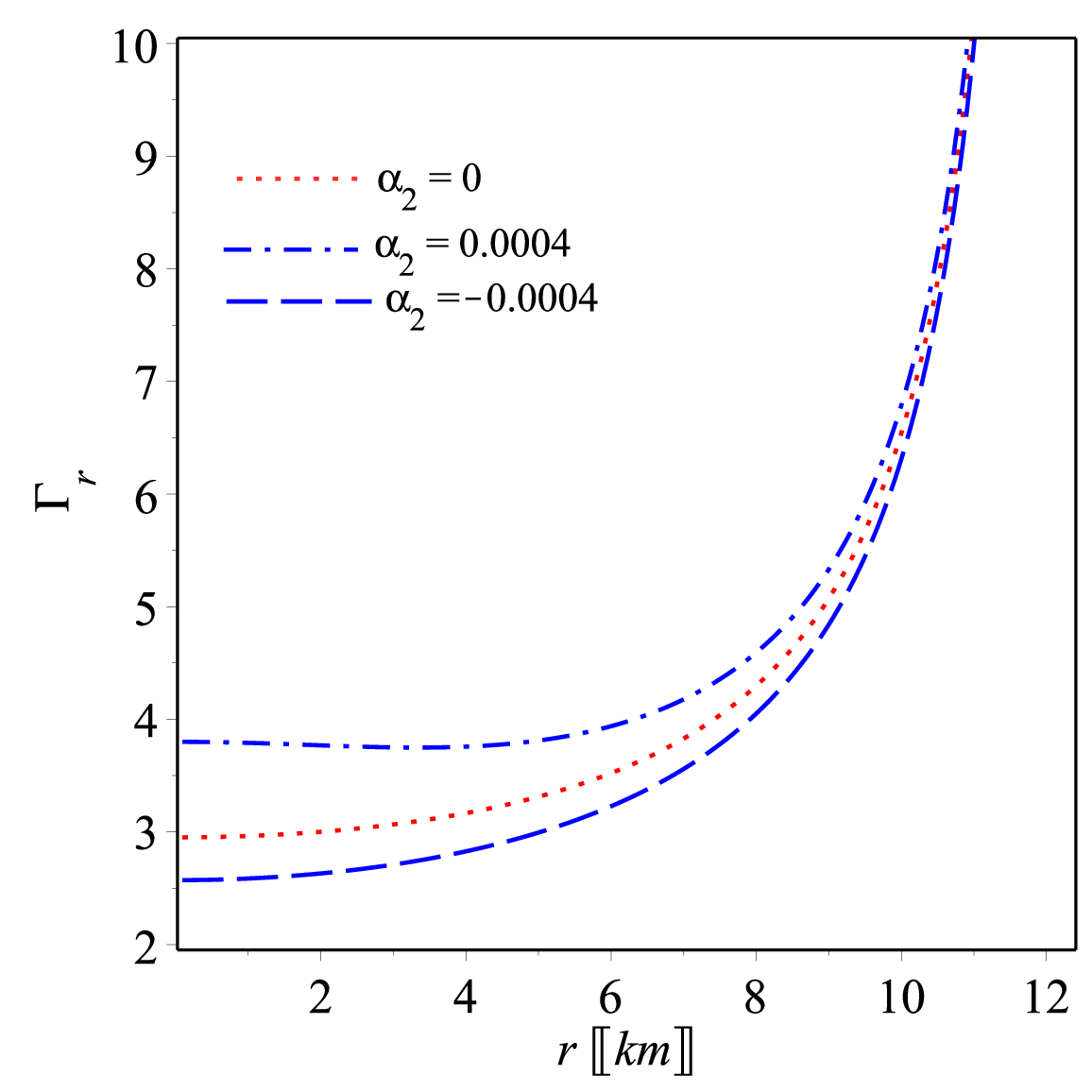}}\hspace{0.2cm}
\subfigure[~$\Gamma_t$]{\label{fig:gamar1}\includegraphics[scale=0.28]{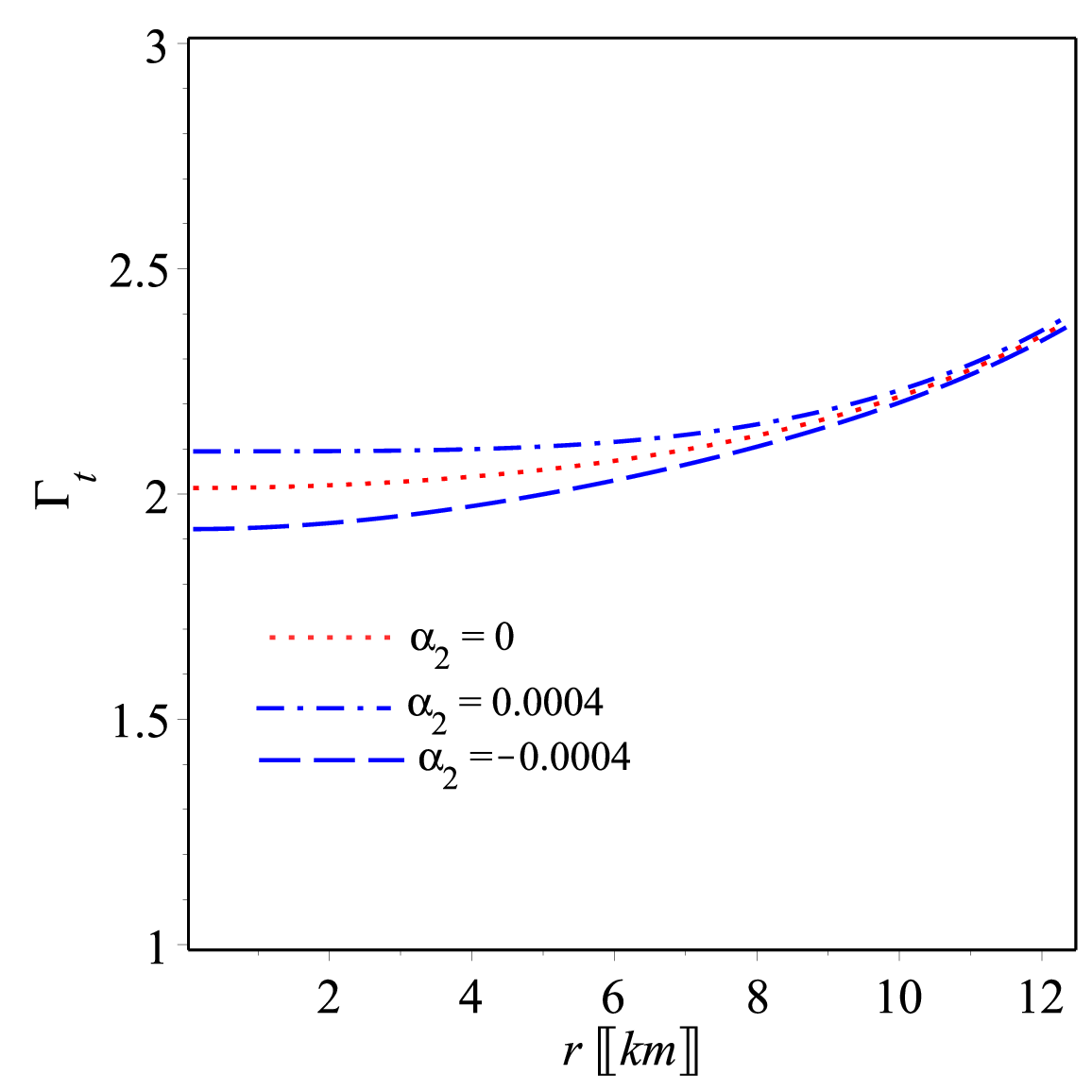}}
\caption{The adiabatic indices, as presented in Eq.~\eqref{eq:adiabatic}, for  ${\mathcal J0740+6620}$.}
\label{Fig:Adiab}
\end{figure}

In this section we are going to use another tool to test the stability of the model under consideration For this aim, we define the following quantities  \citep{Chandrasekhar:1964zz,chan1993dynamical}:
\begin{equation}\label{eq:adiabatic}
{\gamma}=\frac{4}{3}\left(1+\frac{{ \Delta}}{r|{  p}'_r|}\right)_{max},\qquad \qquad
{\Gamma_r}=\frac{{\rho c^2}+{p_r}}{{p_r}}{v_r^2}, \qquad \qquad
{\Gamma_t}=\frac{{\rho c^2}+{p_t}}{{p_t}}{v_t^2, }
\end{equation}
where ${\gamma}$, define the adiabatic index and ${\Gamma_r}$ and ${\Gamma_t}$ are the adiabatic indices in the radial and tangential direction.
Certainly, in the isotropic case where $\Delta=0$, we have $\gamma=4/3$. In the case of mild anisotropy where $\Delta<0$, we find $\gamma<4/3$, similar to what is observed in Newtonian theory. However, in the case of strong anisotropy, we have $\gamma>4/3$. In neutral equilibrium  $\Gamma=\gamma$, while $\Gamma\gamma$  to achieve  a stable equilibrium   \cite{chan1993dynamical,1975A&A....38...51H}. By leveraging Eqs.~\eqref{sol}  we can determine \eqref{eq:pr_grad}--\eqref{eq:pt_grad}, and show that $f(\mathcal{G})$ in its quadratic form offers a stable model for pulsar ${\mathcal J0740+6620}$, regardless of the specific values of $\alpha_2$, as shown in Fig. \ref{Fig:Adiab}. { As is clear from Fig. \ref{Fig:Adiab} \subref{fig:gamar} that the radial pressure diverges at the stellar surface because our model is self bound \cite{Heiselberg:2000dn,Lattimer:2006xb,Lynn:1989xb}}.

\begin{figure}
\centering
\subfigure[~The TOV limitations of GR]{\label{fig:GR}\includegraphics[scale=0.28]{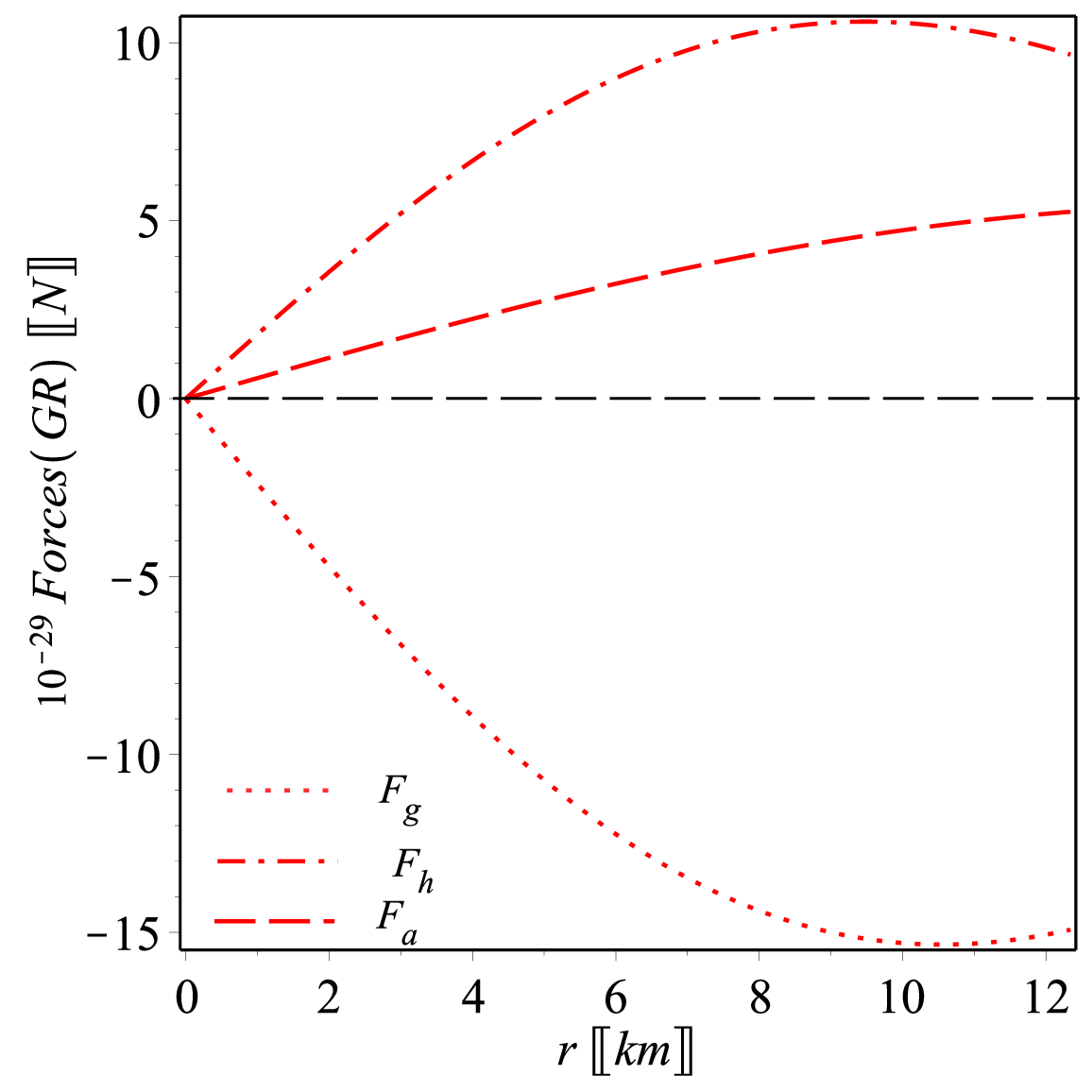}}\hspace{0.2cm}
\subfigure[~The TOV limitations for $\alpha_2>0$]{\label{fig:ve}\includegraphics[scale=0.28]{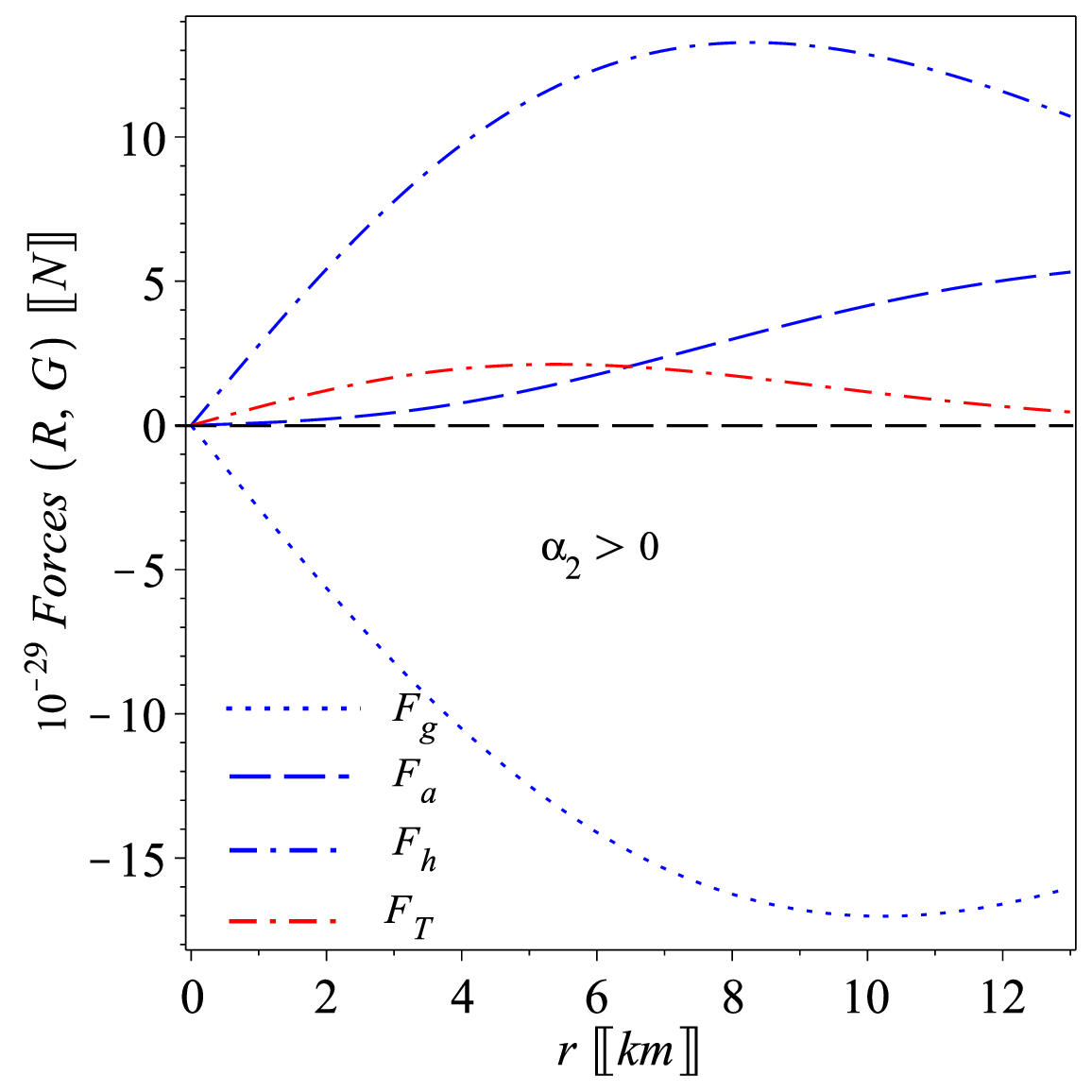}}\hspace{0.2cm}
\subfigure[~The TOV limitations for $\alpha_2<0$)]{\label{fig:nv}\includegraphics[scale=0.28]{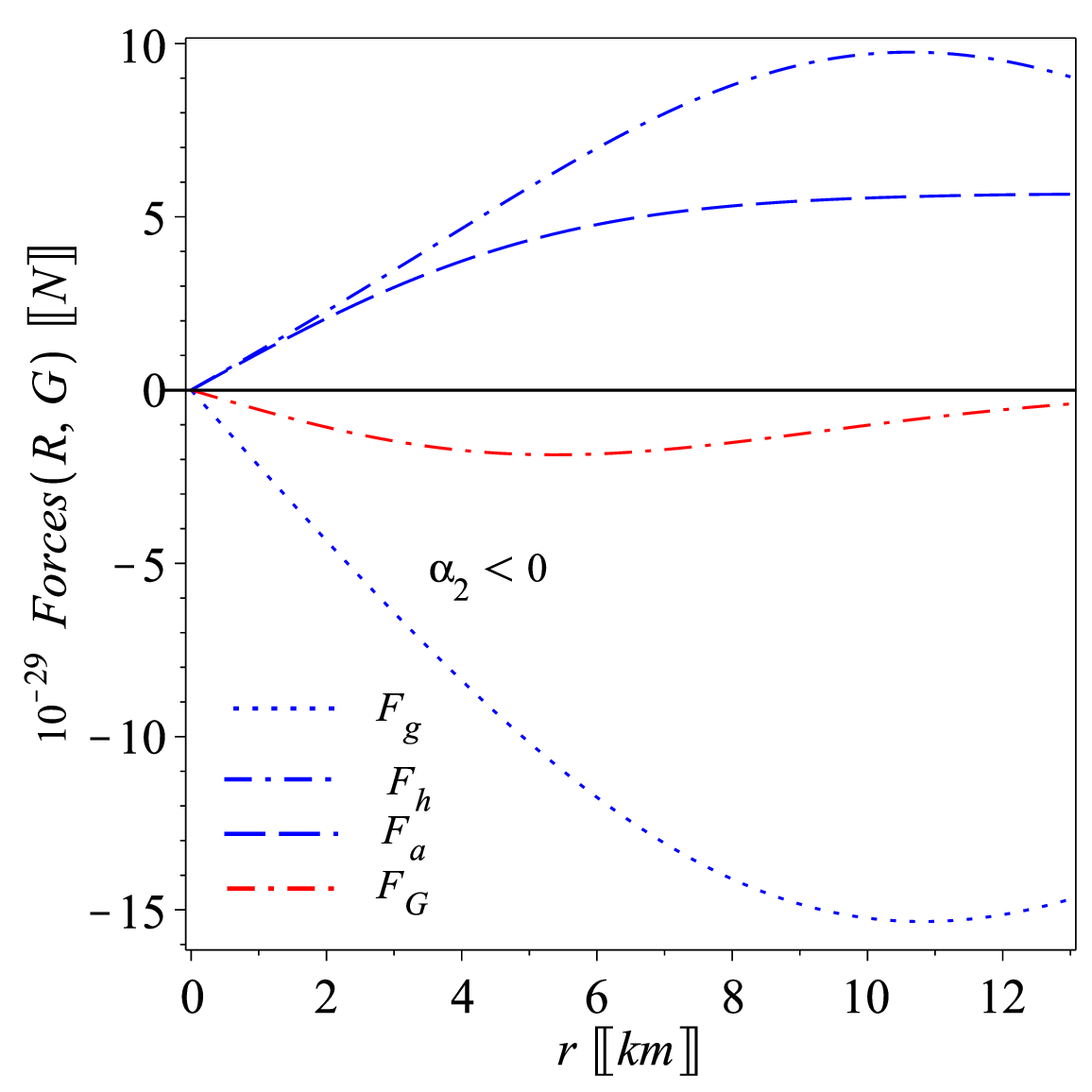}}
\caption{TOV constraint \eqref{eq:RS_TOV}: Various forces, given by Eqs. \eqref{eq:Forces}, acting within the pulsar ${\mathcal J0740+6620}$ are depicted for $\alpha_2=0,~\pm 0.0004$. In the case  $\alpha_2=0.0004$  an extra negative force, further reinforcing the force of gravitational collapse. When $\alpha_2=-0.0004$,  introduces a l positive additional force, that mitigates the collapsing force.}
\label{Fig:TOV}
\end{figure}

{ Next, we explore the hydrostatic equilibrium in the current model using the TOV equation. This equation has been modified to be applicable to a specific context of $\mathcal{R}+f(\mathcal{G})$ theory, and it takes the following form \cite{Tolman:1934za,tolman1939static,oppenheimer1939massive}:}\footnote{ It is of interest to stress that the TOV equation given by Eq. (\ref{eq:RS_TOV}) is different from the original one due to the extra term given by ${\mathcal F_G}$. The appearance of this term is due to the non-conservation given by Eq. (\ref{a1m1}).}
\begin{equation}\label{eq:RS_TOV}
{\mathcal F_a}+{\mathcal F_g}+{\mathcal F_h}+{\mathcal F_G}=0\,.
\end{equation}
in this context, ${\mathcal F_h}$,  ${\mathcal F_a}$, and  ${\mathcal F_g}$, denote hydrostatic,  anisotropic, and gravitational forces, along with the supplementary force ${\mathcal F_\mathcal{G}}$ arising from the effects of the quadratic $\mathcal{R}+f(\mathcal{G})$ gravity theory $f(\mathcal{G})$. Such forces are characterized as:
\begin{eqnarray}\label{eq:Forces}
  {\mathcal  F_a} &=&\frac{ 2{\mathit  \Delta}}{\mathit r} ,\qquad
 {\mathcal  F_g} = -\frac{{\mathit  M_g}}{r}({\mathit  \rho c^2}+{\mathit p_r})e^{\varrho/2} ,\nonumber\\
  {\mathcal  F_h} &=&-{\mathit  p'_t} ,\qquad
 {\mathcal F_\mathcal{G}}=-2\alpha_2(\mathcal{G}').
\end{eqnarray}
Within the force equation for ${\mathcal F_g}$, we present $\varrho:=\psi-\lambda$, and  $M_g$  depicts, in three dimensions, the mass of an isolated system. This quantity is  defined  through the use of the expression presented in  \citep{1930PhRv...35..896T} adapted to the context of $\mathcal{R}+f(\mathcal{G})$ gravity
\begin{eqnarray}\label{eq:grav_mass}
{\mathrm M_g(r)}&=&{\int_{\mathit V}}\Big(\mathfrak{{{ T_1}}}{^r}{_r}+\mathfrak{{{ T_1}}}{^\theta}{_\theta}+{\mathfrak{T_1}}{^\phi}{_\phi}-{\mathfrak{T_1}}{^t}{_t}\Big)\sqrt{-g}\,dV\nonumber\\
&=&e^{-\psi}(e^{\psi/2})'  e^{\lambda/2} r =\frac{1}{2} r \psi' e^{-\varrho/2}\,.
\end{eqnarray}
As a result, the gravitational force is expressed as ${\mathit F_g}=-\frac{n_0 r}{R_s^2}({\mathit \rho c^2}+{ p_r})$. By employing Eqs.~\eqref{sol} and  \eqref{eq:pr_grad}--\eqref{eq:pt_grad}, we can verify Eq.~\eqref{eq:RS_TOV} which ensure by plot \ref{Fig:TOV}.
\section{The relationship between mass and radius and the equation of state}\label{Sec:EoS_MR}

\begin{figure}[th!]
\centering
\subfigure[~ EoS in the radial direction]{\label{fig:RfEoSp}\includegraphics[scale=0.45]{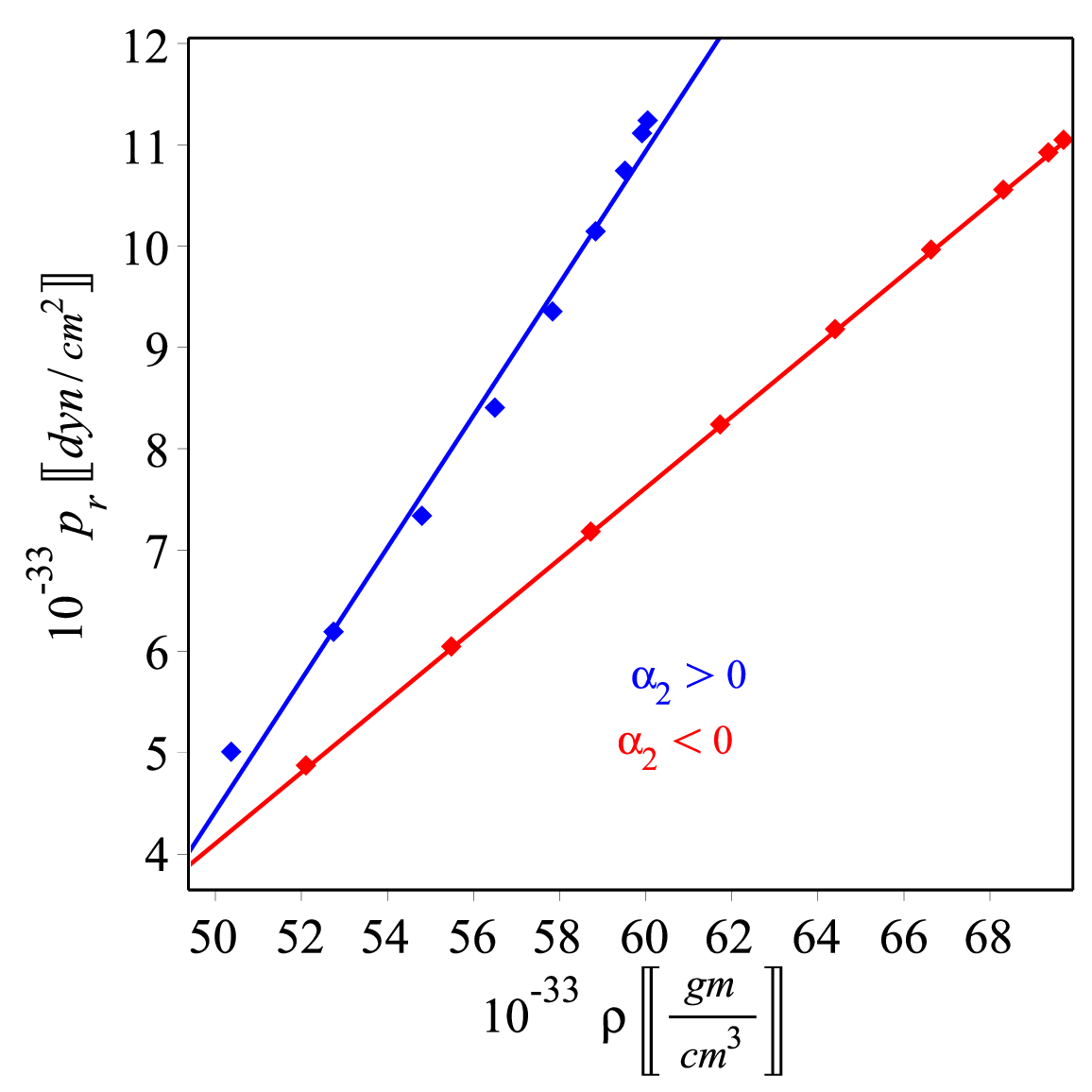}}
\subfigure[~EoS in the  tangential direction]{\label{fig:TEoSn}\includegraphics[scale=0.45]{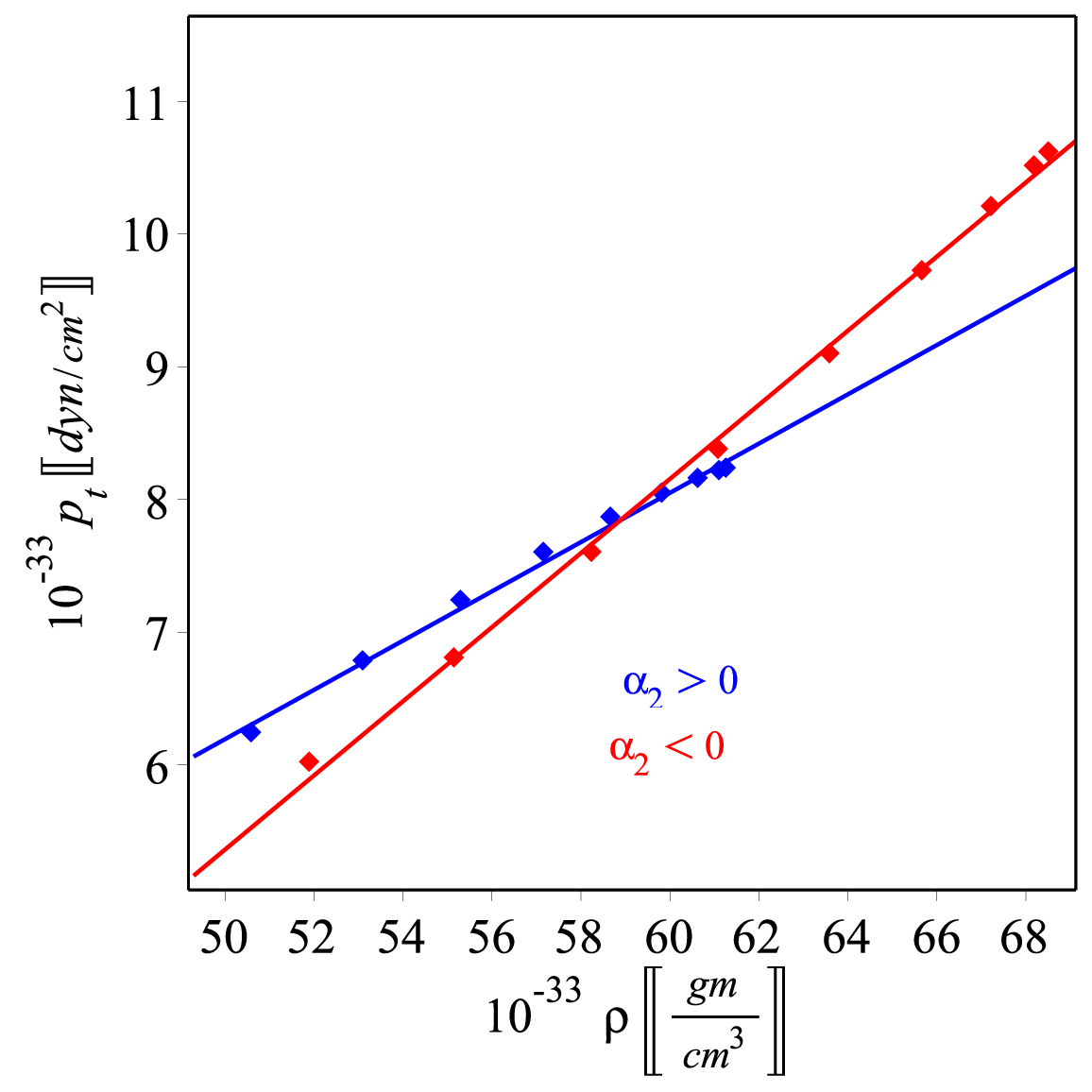}}
\caption{Optimal fit EoSs for the pulsar ${\mathcal J0740+6620}$, as shown in \subref{fig:RfEoSp}, involve the generation of a series of data   for $\rho$ and $p_r$ using Eqs. \eqref{sol} with $\alpha_2=\pm 0.0004$. These observations are consistent with a linear EoS behavior (\subref{fig:TEoSn}. Likewise, when considering tangential EoS's when $\alpha_2=\pm 0.0004$  The data points exhibit robust alignment with a linear trend and align with previously established results, particularly Eq.  \eqref{eq:KB_EoS2}. This reaffirms the validity of these relationships within the pulsar's interior.}
\label{Fig:EoS}
\end{figure}

Astrophysicists are still grappling with the enigmatic composition of matter within the cores of neutron stars, as these cores achieve densities greater than ($\rho_\text{nuc}$), an unexplored domain lying beyond the grasp of earthly laboratories. In spite that the EoS controlling the matter inside neutron stars is still unknown, measurements of their mass and radius in astrophysical settings may be able to constrain it. As a result, astrophysical observations can narrow down the mass-radius diagram linked to a particular  EoS. Indeed, in this study, we do not enforce specific  EoSs; instead, we employ the KB ansatz \eqref{eq:KB}. However,  EoS derived from this ansatz shows a relationship between $\rho$ and pressure (\eqref{eq:KB_EoS2}), which are mainly useful at the core because of their dependence on power series.
 Utilizing the numerical values provided in Section \ref{Sec:obs_const} for the pulsar ${\mathcal J0740+6620}$ and the equations of motion of  $\mathcal{R}+f(\mathcal{G})$  given by  Eqs. \eqref{sol}, we create  Fig. \ref{Fig:EoS}.
%

%
\begin{figure*}[t]
\subfigure[~Compactness versus radius ]{\label{fig:Comp}\includegraphics[scale=0.4]{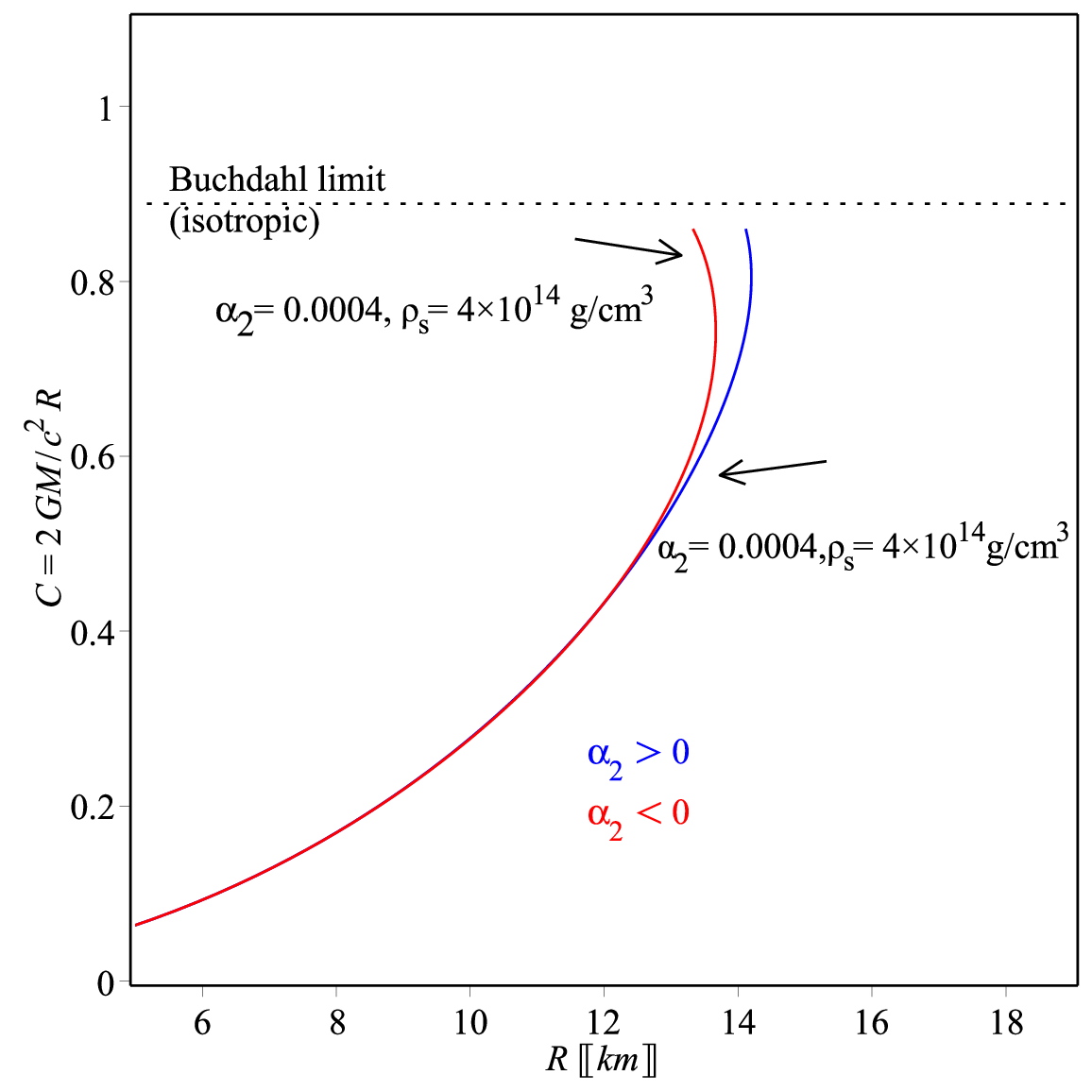}}\hspace{0.5cm}
\subfigure[~Diagram depicting mass as a function of radius]{\label{fig:MR}\includegraphics[scale=.4]{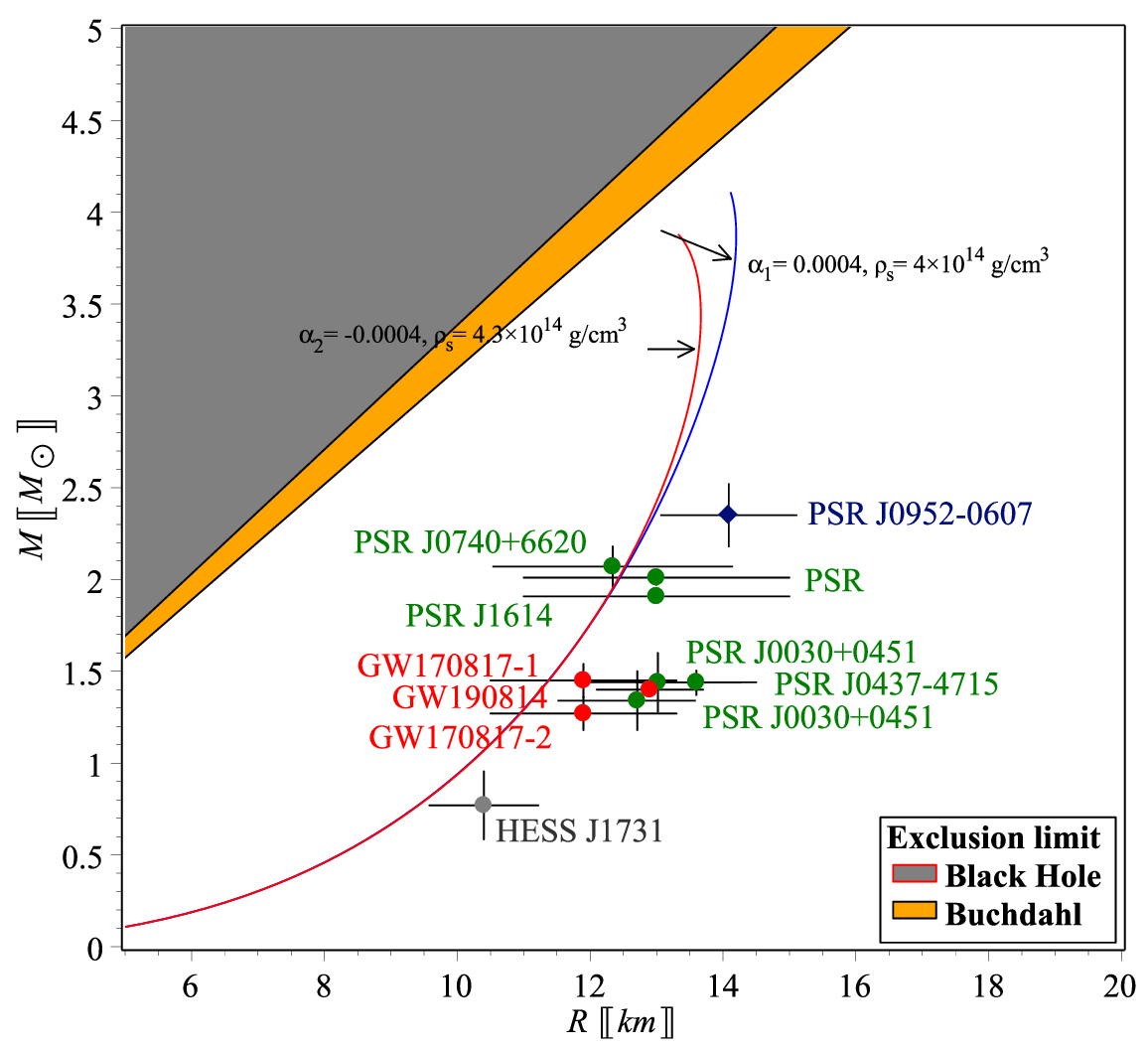}}
\caption{\subref{fig:Comp}. Compactness versus radius plot: The dashed horizontal line denotes the BL at a compactness of $C=8/9$. We illustrate the Compactness-radius curves associated with the optimal EoS fits as provided in Fig. \ref{Fig:EoS}. Furthermore, we depict Compactness-radius curves for  $\rho_s=4\times 10^{14}$ g/cm$^3$. It's worth mentioning that for $\alpha_2=0.0004$ the compactness can be extended up to the BL of $C\to8/9$ a characteristic shared with the  GR  model  \citep{Roupas:2020mvs}. { It is noteworthy that, intriguingly, when $\alpha_2=0.0004$ the extreme value of the compactness is below the BL  for $\rho_s=4\times 10^{14}$ g/cm$^3$. The Mass-radius (MR) plot presented in \subref{fig:MR}, the colors gray (orange) area denotes the black hole limit. Likewise, in the case of positive/negative $\alpha_2$ even at their maximum mass, the mass-radius curves do not cross the BL.}}
\label{Fig:CompMR}
\end{figure*}
%


In this study, concerning $J0740+662$, we determine the BL as follows: For $\alpha_2=\pm0.0004$ we obtain $C\lesssim 0.86$ which results in slight modifications to the GR BL.
  This aligns with our observation that quadratic gravity in this scenario introduces an additional force to TOV equation. This opposing force counteracts gravity, allowing the pulsar to have greater mass and attain greater compactness.
  We assume a surface density of $\rho_\text{s}=4\times 10^{14}$ g/cm$^{3}$ for both $\alpha_2=\pm0.0004$. For various values of the compactness parameter in the range of $0 \leq C \leq 1$ we compute the density profile using Eq. \eqref{sol} to derive the corresponding radius $R$. Likewise, we derive compactness-radius curves, as depicted in plot \ref{Fig:CompMR}\subref{fig:Comp}.

For the previously derived best-fit equations of state, MR curves are provided in plot  \ref{Fig:CompMR}\subref{fig:MR} for positive and negative values of $\alpha_2$.

\section{Conclusion and Discussion }\label{Sec:Conclusion}

The Gauss-Bonnet term holds significant importance in the realms of differential geometry and theoretical physics. It plays a significant role in modern theoretical physics, particularly in the framework of gravitational theories in the frame of astrophysics and cosmology. The Gauss-Bonnet term is a topological invariant, and its quadratic form is often considered in higher-dimensional theories of gravity. In four spacetime dimensions, the Gauss-Bonnet term is a topological invariant and does not contribute to the equations of motion. However, in higher dimensions, it can significantly impact the dynamics of spacetime and matter fields.

Pulsars are crucial astronomical objects, serving as powerful tools for testing various physical theories, including general relativity and alternative theories of gravity. Understanding the behavior of pulsars in the presence of exotic terms in the Einstein-Hilbert action, such as the function $f(\mathcal{G})$ Gauss-Bonnet term, can shed light on the fundamental physics governing these astrophysical phenomena.

In this current study, we explored the quadratic $\mathcal{R}+f(\mathcal{G})$ gravity theory by subjecting it to scrutiny through the lens of astrophysical observations. We examined a plausible situation involving an anisotropic fluid, as is typical for the extremely dense material found within pulsars. Additionally, we supposed that the interior abide to the KB ansatz, which guarantees the smoothness of the inner geometry. Specifically, we leveraged precise measurements of $M$ and $R$ of  ${\mathcal PSR J0740+6620}$, which were determined as $M=2.07 \pm 0.11 M_\odot$ and a radius of $R_s=12.34^{+1.89}_{-1.67}$ km, based on NICER+XMM observations \citep{Legred:2021hdx}. These observations allowed us to delimit $\alpha_1$ of $f(\mathcal{G})$ gravity, i.e. [$\alpha_1$, $C$]. Conversely, ${\mathcal PSR J0740+6620 }$ stands out as one of the most massive pulsars, rendering it an ideal candidate for assessing modified gravity theories.

We have shown the MR plot linked to the best-fit EoS, as shown in Fig. \ref{Fig:CompMR}\subref{fig:MR}. When  $\alpha_2=0.0004$ and with a surface density $\rho_s=4\times 10^{14}$ g/cm$^{3}$ we obtained   $M=4.71 M_\odot$  and $R_s=12.6$ km. When $\alpha_2=-0.0004$ and  $\rho_s=4.0\times 10^{14}$ g/cm$^{3}$, we obtained   $M=4.1 M_\odot$ and $R_s=14.4$ km. These values align perfectly with those observed in the pulsar ${\mathcal J0740+6620}$.

{
To summarize our results:\\
i- We have succeeded for the first time in constructing a stellar model within the framework of the quadratic form of $f(G)$.\\
ii- We have shown that our model is satisfactory from different requirements imposed on any physically admissible
stellar model.\\
iii- Finally, we derive the mass-radius diagram showing the capability of the obtained curves to match observational data of other pulsars.}

\section*{Acknowledgments}
The work of KB was partially supported by the JSPS KAKENHI Grant
Number 21K03547 and 23KF0008.

\appendix A
\section{The KB model and the induced EoSs}\label{Sec:App_1}

It has been shown that the KB ansatz relates the pressures and the density which effectively induces the EoSs as given by Eqs. \eqref{eq:KB_EoS}. The coefficients in those equations are related to the model parameters as listed as below.
\begin{align}
&c_1=\frac{1}{{R_s}^{4}{n_2}^{2}}\left\{ { 4\left( 256\alpha_1{n_0}^{3}n_2 {R_s}^{4}-1376\alpha_1{n_0}^{2}{n_2}^{2}{R_s}^{4}-n_2{R_s}^{4}+2n_0{R_s}^{4} \right) n_2}+\left( \frac{7}2{n_2}^{2}{R_s}^{4}+192\alpha_1{n_0}^{3}{n_2}^{2}{R_s}^{4}-6n_0n_2{R_s}^{4}\right.\right.\nonumber\\
&\left.\left.-{\frac {2464}{3}} \alpha_1{n_0}^{2}{n_2}^{3}{R_s}^{4} \right)  \right\}  \left( 54144 \alpha_1{n_0}^{2}n_2-501888\alpha_1{n_2}^{2}n_0+90+6912\alpha_1\,{n_0}^{3} \right) {c}^{2}\left( 9024\alpha_1{n_0}^{2}n_2-83648\alpha_1 {n_2}^{2}n_0+15\right.\nonumber\\
&\left. +1152\,\alpha_1\,{n_0}^{3} \right) ^{-2},\label{eq:A1}\\[5pt]
&c_2={\frac {(256\alpha_1{n_0}^{3}n_2-1376\alpha_1{n_0}^{2}{n_2}^{2}-n_2+2n_0)}{\kappa{R_s}^{2}}}+ \frac{1}{{R_s}^{2}{n_2}^{2}}\left\{{\frac { 4\left( 256 \alpha_1{n_0}^{3}n_2-1376\alpha_1{n_0}^{2}{n_2}^{2}-n_2+2n_0\right) n_2}{\kappa}}\right.\nonumber\\
&\left.+ \frac{1} {\kappa}\left( \frac{7}2\,{n_2}^{2}+192\, \alpha_1\,{n_0}^{3}{n_2}^{2}-6\,n_0\,n_2-{\frac {2464}{3}}\,\alpha_1\,{n_0}^{2}{n_2}^{ 3} \right) \right\}  \left(54144\,\alpha_1\,{n_0}^{2}n_2 -501888\,\alpha_1 \,{n_2}^{2}n_0+ 90+6912\,\alpha_1\,{n_0}^{3} \right) \nonumber\\
& \left( 2496\,\alpha_1\,{n_2}^{3}n_0-3\,n_2+ 96\,\alpha_1\,{n_0 }^{2}{n_2}^{2} \right)  \left( 9024\,\alpha_1\,{n_0}^{2}n_2-83648\,\alpha_1\,{n_2}^{2}n_0+15+ 1152\,\alpha_1\,{n_0}^{3} \right) ^{-2}
,\label{eq:A2}\\[5pt]
&c_3={\frac {  \left( 18432\,\alpha_1\,{n_0}^{3}\, {R_s}^{2}{c}^{2}-6\,{R_s}^{2}{c}^{2} \right) {n_2}^{2}-68416\,\alpha_1\,{n_0}^{2}\,{R_s}^{2}{c} ^{2}{n_2}^{3}+ \left(18\,n_0\,\kappa\,{R_s}^{2}{c}^{2} -768\,\alpha_1\,{n_0}^{4}{R_s}^{2}{c}^{2} \right) n_2-6\,{n_0}^{2} {R_s}^{2}{c}^{2} }{{R_s}^{2} \left( 9024\,\alpha_1\,{n_0}^{2}n_2-83648\,\alpha_1\,{n_2}^{2}n_0+15+1152\,\alpha_1\,{n_0}^{3} \right) {n_2}^{2}}},\label{eq:A3}\\[5pt]
&c_4=\frac {1}{n_2\,{R_s}^{2}\kappa}  \left\{68672\,{n_2}^{4}\alpha_1\,n_0-55666688\,{n_2}^{5}{\alpha_1}^{2}{n_0}^{3}+18743296\,{n_2}^{4}{\alpha_1}^{2}{n_0}^{4}+53792\,\alpha_1\,{n_0}^{2}{n_2}^{3}
-2961408\,{n_2}^{3}{\alpha_1}^{2}{n_0}^{5}+3 \,{n_2}^{2}-\right.\nonumber\\
&\left.51264\alpha_1{n_0}^{3}{n_2}^{2}+ 368640{n_2}^{2}{\alpha_1}^{2}{n_0}^{6}-24n_0n_2+5184n_2\alpha_1{n_0}^{4}+18{n_0}^{2 }\right\}\left( 9024\alpha_1{n_0}^{2}n_2-83648\alpha_1{n_2}^{2}n_0+15+1152\alpha_1 {n_0}^{3} \right)^{-1}.\label{eq:A4}
\end{align}
Using the above set of equations, one can find the physical quantities appear in Eq. \eqref{eq:KB_EoS2} in terms of the model parameters, where $v_r^2=c_1$, $\rho_1=\rho_s=-c_2/c_1$, $v_t^2=c_3$ and $\rho_2=-c_4/c_3$.

\section{Density and pressures gradients}\label{Sec:App_2}
Recalling the matter density and the pressures as obtained for the quadratic polynomial $f(R)$ gravity, namely Eqs. \eqref{sol}, we obtain the gradients of these quantities with respect to the radial distance as below
\begin{align}\label{eq:dens_grad}
&\rho'=\frac{-2\,{e^{-4{\frac {n_2\,{r}^{2}}{{R_s}^{2}}}}}}{{R_s}^{4}{ \kappa^2}{r}^{7}{c}^{2}} \left( 256{ r}^{6}\alpha_2{n_0}^{2}{n_2}^{2}{R_s}^{2}+256{r}^{6}\alpha_2\,{n_0}^{3}n_2\,{R_s}^{2}+192\,{r}^{4}\alpha_2 \,{n_0}^{2}n_2{R_s}^{4}-3840\,{r}^{2}\alpha_2{R_s}^{6}n_0n_2-10240\,{r}^{6}\alpha_2{n_2}^{3}n_0\, {R_s}^{2}\right.\nonumber\\
&\left.-7168\,{r}^{4}\alpha_2\,{n_2}^{2}n_0\,{R_s}^{4}-192 \,{r}^{8}\alpha_2\,{n_0}^{4}n_2\,{e^{{\frac {n_2\,{r}^{2}}{{R_s}^{2}}}}}-13440\,{r}^{8}\alpha_2\,{n_2}^{4}n_0\,{e^{{\frac {n_2\,{r}^{2}}{{R_s}^{2}}}}}+5184\,{r}^{8 }\alpha_2\,{n_0}^{2}{n_2}^{3}{e^{{\frac {n_2 \,{r}^{2}}{{R_s}^{2}}}}}+768\,{r}^{8}\alpha_2\,{n_0}^{3}{n_2}^{2}{e^{{\frac {n_2\,{r}^{2}}{{R_s}^{2}}}}}\right.\nonumber\\
&\left.-128\,{r}^{2}\alpha_2\,{R_s}^{6}{n_0}^{2}{e^{{\frac {n_2\,{r}^{2}} {{R_s}^{2}}}}}-128\,{r}^{4}\alpha_2\,{a_0}^{3}{e^{{\frac {n_2\,{r}^{2}}{{R_s}^{2}}}}}{R_s}^{4}+64\,{r}^{8}\alpha_2\,{n_0}^{4}{e^{2\,{\frac {n_2\,{r}^{2}}{{R_s}^{2}}}}}n_2+ 768\,{r}^{8}\alpha_2\,{n_2}^{4}n_0\,{e^{2\,{\frac {n_2\,{r}^{2}}{{R_s}^{2}}}}}-704\,{r}^{8}\alpha_2\,{n_0}^{ 2}{n_2}^{3}{e^{2\,{\frac {n_2\,{r}^{2}}{{R_s}^{2}}}}}\right.\nonumber\\
&\left.- 128\,{r}^{8}\alpha_2\,{n_0}^{3}{n_2}^{2}{e^{2\,{ \frac {n_2\,{r}^{2}}{{R_s}^{2}}}}}+64\,{r}^{2}\alpha_2\,{R_s}^{6} {n_0}^{2}{e^{2\,{\frac {n_2\,{r}^{2}}{{R_s}^{2}}}}}+64\, {r}^{4}\alpha_2\,{n_0}^{3}{e^{2\,{\frac {n_2\,{r}^ {2}}{{R_s}^{2}}}}}{R_s}^{4}+{r}^{6}{e^{3\,{\frac {n_2\,{r}^{2}} {{R_s}^{2}}}}}n_2\,{R_s}^{2}+{r}^{4}{e^{3\,{\frac {n_2\,{r }^{2}}{{R_s}^{2}}}}}{R_s}^{4}-{r}^{4}{e^{4\,{\frac {n_2\,{r}^{2 }}{{R_s}^{2}}}}}{R_s}^{4}\right.\nonumber\\
&\left.-1152\,{R_s}^{8}\alpha_2\,n_0\,{e^{2 \,{\frac {n_2\,{r}^{2}}{{R_s}^{2}}}}}+64\,{r}^{4}\alpha_2\,{n_0}^{3}{R_s}^{4}+64\,{r}^{2}\alpha_2\,{R_s}^{6}{n_0}^{2}+ 21504\,{n_2}^{4}{r}^{8}\alpha_2\,n_0-6016\,{n_2}^{3 }{r}^{8}\alpha_2\,{n_0}^{2}-768\,{n_2}^{2}{r}^{8}\alpha_2\,{n_0}^{3}\right.\nonumber\\
&\left.+128\,n_2\,{r}^{8}\alpha_2\,{n_0} ^{4}+2304\,{R_s}^{8}\alpha_2\,n_0\,{e^{{\frac {n_2\, {r}^{2}}{{R_s}^{2}}}}}-2\,{r}^{8}{n_2}^{2}{e^{3\,{\frac {n_2\,{r}^{2}}{{R_s}^{2}}}}}-1152\,{R_s}^{8}\alpha_2\,n_0+5632 \,{r}^{2}\alpha_2\,{R_s}^{6}n_0\,n_2\,{e^{{\frac {n_2\,{r}^{2}}{{R_s}^{2}}}}}\right.\nonumber\\
&\left.+7808\,{r}^{4}\alpha_2\,{R_s}^{4}n_0\,{n_2}^{2}{e^{{\frac {n_2\,{r}^{2}}{{R_s}^{2}}}}}- 320\,{r}^{4}\alpha_2\,{n_0}^{2}n_2\,{e^{{\frac {n_2\,{r}^{2}}{{R_s}^{2}}}}}{R_s}^{4}-384\,{r}^{6}\alpha_2\,{n_0}^{3}n_2\,{e^{{\frac {n_2\,{r}^{2}}{{R_s}^{2}}}}}{R_s}^ {2}-384\,{r}^{6}\alpha_2\,{n_0}^{2}{n_2}^{2}{e^{{ \frac {n_2\,{r}^{2}}{{R_s}^{2}}}}}{R_s}^{2}\right.\nonumber\\
&\left.+8448\,{r}^{6}\alpha_2 \,{n_2}^{3}n_0\,{e^{{\frac {n_2\,{r}^{2}}{{R_s}^{2} }}}}{R_s}^{2}-1792\,{r}^{2}\alpha_2\,{R_s}^{6}n_0\,{e^{2\,{ \frac {n_2\,{r}^{2}}{{R_s}^{2}}}}}n_2-1536\,{r}^{4}\alpha_2\,{R_s}^{4}n_0\,{n_2}^{2}{e^{2\,{\frac {n_2\,{r}^{ 2}}{{R_s}^{2}}}}}+128\,{r}^{4}\alpha_2\,{n_0}^{2}n_2\,{ e^{2\,{\frac {n_2\,{r}^{2}}{{R_s}^{2}}}}}{R_s}^{4}\right.\nonumber\\
&\left.+128\,{r}^{6} \alpha_2\,{n_0}^{3}n_2\,{e^{2\,{\frac {n_2\,{ r}^{2}}{{R_s}^{2}}}}}{R_s}^{2}+128\,{r}^{6}\alpha_2\,{n_0}^{2}{n_2}^{2}{e^{2\,{\frac {n_2\,{r}^{2}}{{R_s}^{2}}}}}{R_s}^{2} -1024\,{r}^{6}\alpha_2\,{n_2}^{3}n_0\,{e^{2\,{ \frac {n_2\,{r}^{2}}{{R_s}^{2}}}}}{R_s}^{2} \right) ,
\end{align}
\begin{align}\label{eq:pr_grad}
  & p'_r=\frac{1}{ {R_s}^{4}{ \kappa^2} {e^{{\frac {4n_2\,{r}^ {2}}{{R_s}^{2}}}}} {r}^{5}}\left( 2\, \left( {e^{{\frac {n_2\,{r}^{2}}{{R_s}^{2}}}}} \right) ^{4}{R_s}^{4}{r}^{2}+ \left( -2\,{R_s}^{4}{r}^{2}-2\,n_2\,{R_s }^{2}{r}^{4}-4\,{r}^{6}n_0\,n_2 \right)  \left( {e^{{ \frac {n_2\,{r}^{2}}{{R_s}^{2}}}}} \right) ^{3}-640\,\alpha_2\, \left( \frac{1}5\,n_2\, \left( n_0+3\,n_2 \right)\right.\right.\nonumber\\
&\left.\left.  \left( n_0-n_2 \right) {r}^{6}+2/5\,{R_s}^{2}n_2\, \left( n_2+n_0 \right) {r}^{4}+1/5\,{R_s}^{4} \left( 6\,n_2+n_0 \right) {r}^{2}+{R_s}^{6} \right) {n_0}^{2} \left( {e^{{ \frac {n_2\,{r}^{2}}{{R_s}^{2}}}}} \right) ^{2}+2304\,\alpha_2 \,{n_0}^{2} \left(\frac{1}6\,n_2\, \left( 6\,n_0\,n_2- 15\,{n_2}^{2}\right.\right.\right.\nonumber\\
&\left.\left.\left.+{n_0}^{2} \right) {r}^{6}+1/3\,{R_s}^{2}n_2 \, \left( n_0+3\,n_2 \right) {r}^{4}+1/9\,{R_s}^{4} \left( n_0+{\frac {33}{2}}\,n_2 \right) {r}^{2}+{R_s}^{6} \right) { e^{{\frac {n_2\,{r}^{2}}{{R_s}^{2}}}}}-1664\, \left( \frac{2}{13}\,n_2\, \left( n_0+21\,n_2 \right)  \left( n_0-3\,n_2 \right) {r}^{6}\right.\right.\nonumber\\
&\left.\left.+{\frac {4}{13}}\,{R_s}^{2}n_2\, \left( 9\,n_2+n_0 \right) {r}^{4}+1/13\,{R_s}^{4} \left( n_0+35\,n_2\right) {r}^{2}+{R_s}^{6} \right) \alpha_2\,{n_0}^{2} \right)\,,
\end{align}
\begin{align}\label{eq:pt_grad}
&p'_t=\frac{1}{{ \kappa^2}  {e^{{\frac {4n_2\,{r}^{2}}{{R_s} ^{2}}}}}{R_s}^{6}{r}^{5}}\left\{ 2\,{r}^{6} \left(  \left( -n_2\,{r}^{2}+{R_s}^{2} \right) { n_0}^{2}+ \left( {n_2}^{2}{r}^{2}-3\,n_2\,{R_s}^{2} \right) n_0+{n_2}^{2}{R_s}^{2} \right)  \left( {e^{{ \frac {n_2\,{r}^{2}}{{R_s}^{2}}}}} \right) ^{3}-128\,{R_s}^{2}{n_0}^{2}\alpha_2\, \left( n_0\,{r}^{2}+{R_s}^{2}\right.\right.\nonumber\\
&\left.\left.-n_2\,{r}^ {2} \right)  \left( {R_s}^{4}+{r}^{2}{R_s}^{2}n_2+n_2\,{r}^{4}n_0-{n_2}^{2}{r}^{4} \right)  \left( {e^{{\frac {n_2\,{r}^{2}}{{R_s}^{2}}}}} \right) ^{2}-768\,\alpha_2\,{n_0}^{ 2} \left(  \left( -1/6\,n_2\,{r}^{6}{R_s}^{2}-{n_2}^{2}{r}^{8} \right) {n_0}^{2}+ \left( 6\,{n_2}^{3}{r}^{8}-3\,{n_2} ^{2}{r}^{6}{R_s}^{2}\right.\right.\right.\nonumber\\
&\left.\left.\left.-2\,n_2\,{r}^{4}{R_s}^{4}-\frac{2}3\,{r}^{2}{R_s}^{6} \right) n_0+{\frac {43}{6}}\,{n_2}^{3}{r}^{6}{R_s}^{2}+{ \frac {23}{6}}\,n_2\,{r}^{2}{R_s}^{6}+7\,{n_2}^{2}{r}^{4}{R_s}^{ 4}-5\,{n_2}^{4}{r}^{8}+{R_s}^{8} \right) {e^{{\frac {n_2 \,{r}^{2}}{{R_s}^{2}}}}}+896\,\alpha_2\,{n_0}^{2} \left( \frac{2}7\,n_2\,{r}^{6} \left( {R_s}^{2}\right.\right.\right.\nonumber\\
&\left.\left.\left.-8\,n_2\,{r}^{2} \right) {n_0 }^{2}-\frac{3}7\,{r}^{2} \left( 4\,n_2\,{R_s}^{4}{r}^{2}+{R_s}^{6}-{\frac { 160}{3}}\,{r}^{6}{n_2}^{3}+{\frac {76}{3}}\,{n_2}^{2}{R_s}^{2} {r}^{4} \right) n_0+{R_s}^{8}-48\,{n_2}^{4}{r}^{8}+{\frac {68} {7}}\,{n_2}^{2}{r}^{4}{R_s}^{4}+{\frac {31}{7}}\,n_2\,{r}^{2}{ R_s}^{6}\right.\right.\nonumber\\
&\left.\left.+{\frac {274}{7}}\,{n_2}^{3}{r}^{6}{R_s}^{2} \right) \right\}\,.
\end{align}


\end{document}